\documentclass[a4paper, twocolumn=off, twoside=on]{scrartcl} 

\KOMAoptions{fontsize=11pt} %
\KOMAoptions{BCOR=0mm}		%
\PassOptionsToPackage{hidelinks}{hyperref}

\usepackage{amsmath}
\usepackage{amsfonts}
\usepackage{siunitx}
\usepackage{float}
\usepackage{caption}
\usepackage{subcaption}
\usepackage{booktabs}
\usepackage[usenames,dvipsnames]{xcolor}
\usepackage[hidelinks]{hyperref}

\usepackage{siunitx}
\usepackage[fixamsmath]{mathtools}
\mathtoolsset{showonlyrefs=true}

\usepackage{enumitem}
\setlist{noitemsep}
\setitemize[2]{label=\textperiodcentered}

\usepackage{etoolbox}

\usepackage{amsthm}
\theoremstyle{plain}
\newtheorem{theorem}{Theorem}[section]

\newtheorem{corollary}[theorem]{Corollary}

\theoremstyle{remark}
\newtheorem{remark}[theorem]{Remark}

\theoremstyle{definition}
\newtheorem{definition}[theorem]{Definition}
\newtheorem{assumption}[theorem]{Assumption}

\usepackage{fontenc}

\usepackage{geometry}
\usepackage[proportional]{libertinus}
\usepackage{libertinust1math}

\usepackage[backend=biber, style=numeric, sortcites=true, citestyle=numeric-comp, giveninits=true, sorting=nyt, doi=true, isbn=false, url=false, maxcitenames=3, maxbibnames=10, minalphanames=3, block=none, date=year]{biblatex}
    \AtEveryBibitem{%
      \clearlist{language}%
      \clearfield{type}%
      \clearfield{pages}%
      \clearfield{pagetotal}%
      \clearfield{file}%
      \clearlist{address}%
      \clearlist{location}%
      \clearlist{venue}%
    }

\renewbibmacro{in:}{}

\KOMAoptions{parskip=half}
\KOMAoptions{abstract=true}

\addtokomafont{section}{\centering}
\addtokomafont{titlehead}{}
\addtokomafont{subject}{\normalfont\sffamily\normalsize}
\addtokomafont{title}{\normalfont\sffamily\normalfont\sffamily\Large}
\addtokomafont{subtitle}{\normalfont\sffamily\large}
\addtokomafont{author}{\normalfont\sffamily\large}
\addtokomafont{date}{\normalfont\sffamily\large}
\addtokomafont{publishers}{\normalfont\sffamily\normalsize}

\usepackage{authblk}

\usepackage[en-US]{datetime2}
\DTMlangsetup{showdayofmonth=false}

\setenumerate[1]{label={\arabic*}., ref={\arabic*},  leftmargin=*}
\setitemize{leftmargin=*}
\setlist{rightmargin=1em}
\setlist{noitemsep}

\setkomafont{pagehead}{\normalfont\footnotesize\itshape}
\setkomafont{pagefoot}{\normalfont\footnotesize}
\setkomafont{pagenumber}{\normalfont\normalsize\itshape}

\usepackage[many]{tcolorbox}
\tcbuselibrary{theorems}
\tcbset{breakable, enhanced jigsaw, 
colback=white, 
colframe=black!15, 
coltitle=black, colbacktitle=black!15, fonttitle=\bfseries\sffamily, arc=0mm,
subtitle style={colback=black!15,fontupper={\sffamily\color{black}}}
}
\tcbset{shield externalize}

\newtcolorbox[use counter*=theorem]{boxalgorithm}[2][]{
list text={#2},
  title={Algorithm \thetcbcounter: #2},
  #1
}

\newcommand{\pliq}{-}
\newcommand{\pvap}{+}
\newcommand{\phasel}{\pliq}
\newcommand{\phaser}{\pvap}

 \newcommand{\facet}{S}

\newcommand{\setsep}{\allowbreak:\allowbreak}

\newcommand{\UU}{\vec{U}} 

\renewcommand{\vec}[1]{\mathbf{#1}}

\newcommand{\bN}{\mathbb{N}}

\newcommand{\bR}{\mathbb{R}}

\newcommand{\cT}{\mathcal{T}}

\newcommand{\cV}{\mathcal{V}}

\newcommand{\vx}{\vec{x}}

\newcommand{\ipnt}{v_{\mathrm{int}}} 
\newcommand{\xmin}{\Delta x_{\mathrm{min}}}
\newcommand{\xminInterface}{\Delta x_{\Gamma,\mathrm{min}}}
\newcommand{\xmaxInterface}{\Delta x_{\Gamma,\mathrm{max}}}

\DeclareMathOperator{\dd}{d}
\DeclareMathOperator{\ddd}{d\!}

\DeclarePairedDelimiter\abs{\lvert}{\rvert}

\DeclarePairedDelimiterX{\norm}[1]{\lVert}{\rVert}{
\ifblank{#1}{\:\cdot\:}{#1}
}

\DeclarePairedDelimiter\ceil{\lceil}{\rceil}

\usepackage[clines, plainfootsepline, automark]{scrlayer-scrpage}
\clearscrheadfoot

\ihead[]{}
\cehead{M. Alk\"amper, J. Magiera, C. Rohde}
\cohead{An Interface Preserving Moving Mesh in Multiple Space Dimensions}
\ohead[]{\pagemark}
\ifoot[]{}
\ofoot[]{}

\title{An Interface Preserving Moving Mesh \\ in Multiple Space Dimensions}
\author{Maria Alk\"{a}mper}
\author{Jim Magiera}
\author{Christian Rohde}
\affil{University of Stuttgart, Institute of Applied Analysis and Numerical Simulation}

\begin{document}

\maketitle

\begin{abstract}
An interface preserving moving mesh algorithm in two or higher dimensions is presented. 
It resolves a moving \((d-1)\)-dimensional manifold directly within the \(d\)-dimensional mesh, which means that the interface is represented by a subset of moving mesh cell-surfaces. 
The underlying mesh is a conforming simplicial partition that fulfills the Delaunay property. 
The local remeshing algorithms allow for strong interface deformations.
We give a proof that the given algorithms preserve the interface after interface deformation and remeshing steps.
Originating from various numerical methods, data is attached cell-wise to the mesh.
After each remeshing operation the interface preserving moving mesh retains valid data by projecting the data to the new mesh cells.\newline
An open source implementation of the moving mesh algorithm is available at \cite{alkaemper:interface:2021}.
\end{abstract}

\section{Introduction}
\label{sec:introduction}

\begin{figure}[tp]
 \centering
 \subcaptionbox{\label{subfig:topology_changes_0}}
 [0.3\linewidth]{
\includegraphics[width=0.90\linewidth]{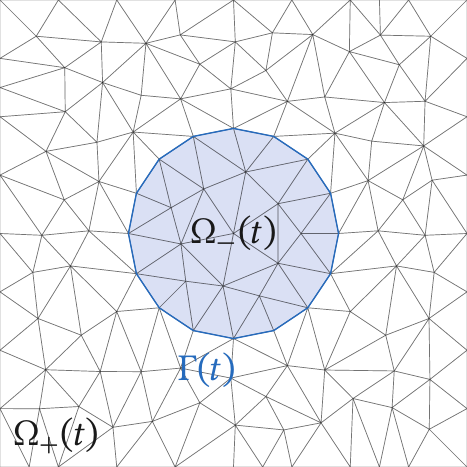}
 }\subcaptionbox{\label{subfig:topology_changes_1}}
 [0.3\linewidth]{
\includegraphics[width=0.90\linewidth]{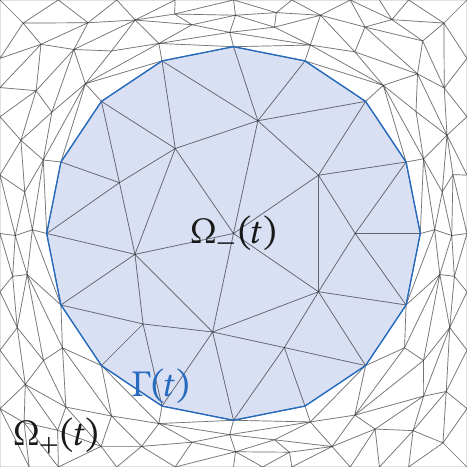}
 }\subcaptionbox{\label{subfig:topology_changes_2}}
 [0.3\linewidth]{
\includegraphics[width=0.90\linewidth]{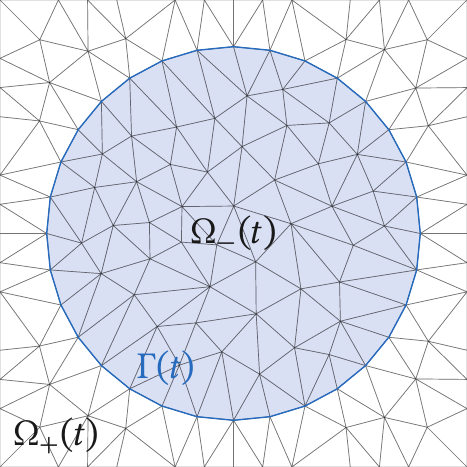}
 }\caption{(\subref{subfig:topology_changes_0}) 
 Initial configuration. 
 (\subref{subfig:topology_changes_1}) 
 Strong interface deformation without and (\subref{subfig:topology_changes_2}) 
 with mesh topology changes.}
 \label{fig:topology_changes_mesh}
\end{figure}

For many applications in science and engineering, mesh-based methods are utilized to solve time-dependent partial differential equations.
The quality of the numerical solutions strongly depends on the quality of the mesh.
This is especially the case for sharp-interface approaches, where the interface should be resolved within the mesh, which means that the interface should coincide with mesh surfaces.
This can be used for example to track discontinuities, such as material faults, the dynamic position of a phase transition in a liquid, or a growing crack in a medium.
\newline
One big challenge of those models is that the interface is free and should be able to move according to the dynamics of the model.
That means the mesh should be able to incorporate large deformations of the interface.
However, a naive implementation by e.g. a Lagrangian approach that does not perform topological changes within the mesh may yield poor meshes --- see Figure~\ref{fig:topology_changes_mesh}.
\newline
To remedy this problem, we propose an
\emph{interface preserving moving mesh (IPMM)}
that has the following advantages:
\begin{itemize}
 \item Strong deformations of the interface are possible.
 \item The interface has an explicit and connected representation within the mesh, from which all geometric information can be inferred, e.g. the local curvature.
 \item A high resolution of the interface is achievable.
 \item Remeshing is performed in a local fashion.
\end{itemize}
The moving mesh ansatz for one space dimension has been proposed in \cite{harten.hyman:self:1983}, and further developed in \cite{sanders:moving:1985}.
A noteworthy contribution to simulate fluid interfaces using front tracking is presented in \cite{unverdi.tryggvason:front:1992,tryggvason.scardovelli.ea:direct:2011},
where fluid interfaces are represented by a separate, lower-dimensional moving mesh that exists on top of a static background mesh.
\newline
The groundwork for our moving mesh approach was laid in \cite{chalons.engel.ea:conservative:2014,chalons.rohde.ea:finite:2017}.
In these contributions hyperbolic conservation laws are discretized using a moving mesh in one and two dimensions, that tracks discontinuous PDE-solutions within the mesh.
\newline
The moving mesh methods applied in \cite{baker.cavallo:dynamic:1999,perot.nallapati:moving:2003,dai.schmidt:adaptive:2005,quan.schmidt:moving:2007,quan.lou.ea:modeling:2009,tukovic.jasak:simulation:2008} 
are related to our work but differ in methodological approaches.
In \cite{perot.nallapati:moving:2003} the authors used mesh motion based on spring analogy in combination with mesh flips in order to maintain optimal mesh connectivity but did not include insertion of removal of points.
Unlike in the latter, in \cite{baker.cavallo:dynamic:1999} insertion and removal of points was included in a mesh enrichment and mesh coarsening step but only for the two-dimensional setting.
Nevertheless, the mesh attained a worse quality after the interface has passed through.
In \cite{dai.schmidt:adaptive:2005,quan.schmidt:moving:2007,quan.lou.ea:modeling:2009,tukovic.jasak:simulation:2008} both the generalization to the three-dimensional case as well as the poor mesh quality was tackled.
The authors presented the insertion and removal of vertices via the edge bisection and edge collision methods as well as the local optimization of the mesh quality via edge swapping and mesh smoothing operations.

In contrast to all aforementioned moving mesh approaches we use a background mesh to retain an initially good mesh quality away from the interface, even after an interface has passed through some part of the mesh.
This is especially beneficial if the initial mesh is a highly optimized (Delaunay) mesh with respect to quality, which can be constructed by well-established methods.
Furthermore, we retain the Delaunay property of the mesh.
Thus, mesh modifications are realized by using built-in operations on Delaunay meshes (insert, remove, find conflicts) only without altering the mesh manually, which gives comparably simple algorithms.
 
We seek to present our moving mesh algorithm in a generic way, to be applicable to a wide range of models, see e.g. \cite{chalons.magiera.ea:finite:2018}, as well as to arbitrary dimensions.
The algorithm is also formulated in a way to be easily implemented in already existing mesh frameworks.
The most important advantages of our moving mesh method become apparent if we regard it in conjunction with the numerical discretization schemes that can be build on top of it (see for example Section~\ref{subsec:mmesh_benchmark:circular}).
One advantage is that no interpolation of the data across the interface is needed --- which could break two-phase flow simulations of compressible fluids.
For example, the interpolated data could be an averaged fluid density over different phases, which cannot be uniquely assigned to a phase and thus would not have a physical meaning.
As another advantage, it is possible to apply an easily exchangeable, separate model to describe the interface dynamics, enabling us to include microscale effects directly at the interface.

In this paper, we will first
introduce our moving mesh ansatz in multiple dimensions by proposing algorithms that allow us to move vertices in the mesh, while taking care of data associated with the mesh (Section~\ref{sec:moving_mesh}).
Subsequently, these methods are used to formulate the IPMM-algorithm in Section~\ref{sec:mmesh:interface_preserving}.
In Section~\ref{sec:mmesh_benchmark} we present computational test cases, showing the capabilities of the method.

To ensure a comprehensible method and reproducible results we provide an open source implementation of the moving mesh algorithm at \cite{alkaemper:interface:2021}.

\section{Moving Mesh Algorithm in Multiple Dimensions}
\label{sec:moving_mesh}

In this section we present the moving mesh approach for \(d \in \{2,3\}\)  dimensions, 
but all algorithms are also applicable for \(d > 3\).
The moving mesh approach is based on simplicial meshes, therefore we begin by defining the basic mesh entities.
\begin{definition}[Simplex, Facet and Edge]
A \(d\)-dimensional \emph{simplex} \(C\) is a polytope defined by  \(d+1\) distinct vertices \(v_0,\ldots,v_{d} \in \bR^d\), i.e.,
\begin{align}
 C \coloneqq \Bigl\{ 
 &
 x \in \bR^d \setsep
 x = \sum\limits_{i=0}^{d} \theta_i v_i, 
 \sum\limits_{i=0}^d \theta_i = 1, 
 \theta_i \geq 0 \text{ for all } i=0,\ldots,d \Bigr\}.
\end{align}
A simplex \(C\) is called \emph{nondegenerate} if the vertices \(v_0,\ldots,v_{d} \in \bR^d\) are not coplanar, i.e., do not lie on a \((d-1)\)-dimensional plane.
\newline
The surface of a \(d\)-dimensional simplex \(C\) consists of \(d+1\) \emph{facets} \(S_l\), \(l = 0,\ldots,d\), given by
\begin{align}
  S_l \coloneqq \Bigl\{ 
  &
  x \in \bR^d \setsep
  x = \sum\limits_{\substack{i=0 \\ i \neq l}}^{d} \theta_i v_i,
   \sum\limits_{\substack{i=0 \\ i \neq l}}^d \theta_i = 1, 
  \theta_i \geq 0 \text{ for all } i=0,\ldots,d,~ i  \neq l \Bigr\}.
\end{align}
Note that a facet is a \((d-1)\)-dimensional simplex in a \(d\)-dimensional space.
An \emph{edge} \(e_{ij}\) of a \(d\)-dimensional simplex, for \(d \geq 2\), is defined by two vertices \(v_i\), \(v_j\), \(i,j = 0,\ldots,d\), \(i\neq j\),
\begin{align}
  e_{ij} \coloneqq  \Bigl\{ 
  &
  x \in \bR^d \setsep
   x = \theta_i v_i + \theta_j v_j 
   \text{ with }
   \theta_i + \theta_j = 1 \text{ and } \theta_i, \theta_j \geq 0 \Bigr\}.
\end{align}
Consequently, a \(d\)-dimensional simplex has \(\tfrac{1}{2}d(d+1)\) edges.
\end{definition}
The simplex surfaces are called \emph{facets} to be consistent with the three-dimensional case.
In two dimensions, both facets and edges correspond to the edges of a triangle. 

To begin we define the static mesh for a given set of vertices \(\cV \subset \bR^d\), which implicitly defines the domain \(\Omega = \operatorname{conv}(\cV)\) by its convex hull.
Naturally, if only a domain \(\Omega \subset \bR^d\) is given, a vertex set \(\cV\) can be chosen whose convex hull is an
approximation of \(\Omega\).
Moreover, we make the following assumption throughout this contribution.
\begin{assumption}
  \label{assumption:general_position}
  We assume that all vertices \(v \in \cV\) are in \emph{general position} \cite{yale:geometry:1968}, i.e., no \(i+1\) vertices lie in a subspace of dimension \(i-1\), for \(i = 1,\ldots, d\), and no vertices are cospherical, which means no \(d+2\) vertices \(v \in \cV\) share the same circumsphere.
\end{assumption}
For most cases Assumption~\ref{assumption:general_position} does not yield severe restrictions but simplifies the presentation of our method by avoiding the discussion of degenerate cases. Still, our method is presented on a sufficiently abstract level, that if all underlying geometric operations are appropriately implemented\footnote{This is for example the case for the  CGAL-framework \cite{project:cgal:2020}, where the developers provide a reliable implementation of basic geometry operations.
See \cite{kettner.mehlhorn.ea:classroom:2008} for some examples that highlight problems that arise by using inexact arithmetic in geometric applications.},
these cases are covered as well.
\begin{definition}[Mesh]
 Let a vertex set \(\cV \subset \bR^d\) be given and let \(\Omega = \operatorname{conv}(\cV)\) be the corresponding domain.
 A \emph{mesh} \(\tau \coloneqq \{C_j \setsep  j = 1, \ldots,N\}\) is a partition of \(\Omega\) into a set of \(N \in \bN\) nondegenerate \(d\)-dimensional simplices \(C_j\), where the vertices of the simplices are elements of \(\cV\).
 The vertices of the \(j\)-th cell \(C_j\) are denoted by \(v^j_i \in \cV\) for \(i=0,\ldots,d\).
\end{definition}
We refer to the simplices also as \emph{cells} in the context of meshes, and the corresponding facets of the mesh are also called \emph{cell surfaces}.

We propose a Delaunay mesh \cite{delaunay:sur:1934,samet:foundations:2006,berg.cheong.ea:computational:2008} as the base of the moving mesh.
\begin{definition}[Delaunay mesh]
 Let a mesh \(\tau\) with vertex set \(\cV\) be given.
 The mesh is called a \emph{Delaunay mesh} (or: \emph{Delaunay triangulation}) iff the circumsphere of each cell \(C_j \in \tau\) of the mesh does not contain any other vertex from \(\cV\) than the cell vertices \(v^j_i\), \(i=0,\ldots,d\).
\end{definition}
Using a Delaunay mesh has several advantages.
It is uniquely defined \cite{delaunay:sur:1934} in the non-cospherical case, and we write in this case
\begin{align}
  \tau = \tau(\cV)
\end{align}
for the  Delaunay mesh of the vertex set \(\cV\).
In case of \(d+2\) cospherical vertices the Delaunay mesh can still be defined, but is not unique anymore.
From a numerical point of view a Delaunay mesh is usually a good choice among the set of all possible triangulations, as it maximizes the minimum angle inside its cells \cite{berg.cheong.ea:computational:2008,boissonnat.yvinec:algorithmic:1998}.
There exist efficient and already implemented algorithms for the insertion and removal of vertices from a Delaunay mesh, see e.g. \cite{boissonnat.devillers.ea:incremental:2009,devillers.teillaud:perturbations:2003}.

\begin{figure}[tp]
 \centering
\includegraphics[width=0.75\linewidth]{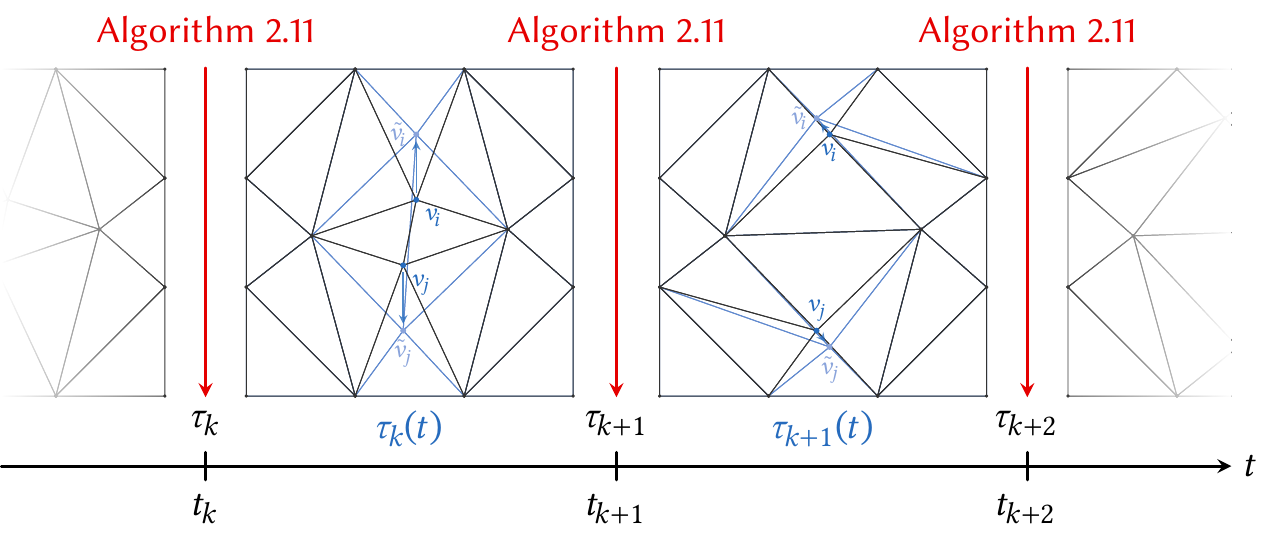}
 \caption{Illustration of a moving mesh in time.}
 \label{fig:remeshing_steps}
\end{figure}

Next, we define the moving mesh introducing time dependence on the mesh.
Hereby, the moving mesh is driven by the motion of the vertices.
Let a monotone sequence of points in time \((t_k)_{k \in \bN_0}\) be given.
For any time interval \([t_k, t_{k+1})\) let a mesh \(\tau_k\) with vertex set \(\cV_k\) be given.
Assume that for each \(v\in \cV_k\) there exists a continuous function
\begin{align*}
 m_v\colon [t_k, t_{k+1}) \to \bR^d, t \mapsto m_v(t),
\end{align*}
with \(m_v(t_k) = \vec{0}\), describing the relative motion of the vertex \(v\).
The position of the vertex \(v\) at time \(t \in [t_k, t_{k+1})\)  is given by \(v(t) \coloneqq v + m_v(t)\).
Then we introduce the moving mesh.
\begin{definition}[Moving Mesh]
 \label{def:moving_mesh}
 If for all \(k \in \bN_0\) and for each \(t \in [t_k, t_{k+1})\), all \emph{moving cells} with vertices \(v^j_i(t) \coloneqq v^j_i + m_{v^j_i}(t)\), defined as
 \begin{align*}
  C_j(t) \coloneqq \Bigl\{
  &
  x \in \bR^d \setsep
  x = \sum\limits_{i=0}^{d} \theta_i (v^j_i + m_{v^j_i}(t)), 
  \sum\limits_{i=0}^d \theta_i = 1, 
  \theta_i \geq 0 \text{ for all } i=0,\ldots,d \Bigr\},
 \end{align*}
 form a mesh of \(\Omega\),
 then we call \(\cT = (\tau_k, \{m_v\}_{v\in \cV_k})_{k \in \bN_0}\) a \emph{moving mesh}.
 The set
 \begin{align*}
 \tau_k(t)
 \coloneqq \{C_j(t) \setsep  C_j \in \tau_k\}, \text{ for } t \in [t_k, t_{k+1}),
 \end{align*}
 is the \emph{moving mesh at time \(t \in [t_k, t_{k+1})\)}, with time dependent vertices \(v(t)\) and cells \(C_j(t)\).
\end{definition}
In general, the Delaunay property of the mesh has to be restored after motion.
Thus, we have to be aware that the topology of the mesh, or more precisely the topology of the underlying graph of the mesh, may change at certain points in time.
That means that due to the motion, the number of cells as well as neighborhood relations may change.
Note that in Definition~\ref{def:moving_mesh}, we have defined the entire moving mesh \(\cT\) as a sequence of base-meshes \(\tau_k\) for every point in time \(t_k\), \(k \in \bN_0\).
Based on these meshes \(\tau_k\), we define time-dependent sub-meshes \(\tau_k(t)\) in the time interval \(t \in [t_k, t_{k+1})\), which have time dependent vertices, but a fixed mesh topology.
Implicitly, this restricts the vertex motion, to avoid degenerate mesh structures.
The mesh topology only changes at each time \(t_{k+1}\) from the base-mesh \(\tau_{k}\) to the next base-mesh \(\tau_{k+1}\), see Figure~\ref{fig:remeshing_steps} for an illustration of a moving mesh.
\newline
The advantage of this approach is that we are able to formulate numerical algorithms on the sub-meshes, where smooth, time-dependent vertex motions are possible, but no topology change takes place.
The necessary topology changes, to allow for e.g. large vertex motions, take place only at discrete points in time \(t_k\), \(k \in \bN_0\).
Consequently, data associated with the mesh has to be transferred only at these discrete points in time,
which will be discussed later in this chapter.

To avoid special treatment of the domain boundary and some edge cases,
we make the assumption that the vertices on the domain boundary
\(\partial \Omega = \partial\operatorname{conv}(\cV)\) do not change their positions, i.e.,
for all \(v\in \cV_k\) with \(v\in\partial \Omega\),
we assume \(m_v \equiv \vec{0}\).

In order to make the moving mesh \(\tau(t)\) unique for a given vertex set \(\cV\) at every discrete point in time \(t_k\), \(k \in \bN_0\), we base the moving mesh on Delaunay meshes. 
Beside their uniqueness, they are generally considered well-shaped meshes for numerical applications. 
\begin{definition}[Delaunay Moving Mesh]
 Let a moving mesh
 \begin{align}
   \cT \allowbreak = \allowbreak (\tau_k, \allowbreak \{m_v\}_{v\in \cV_k})_{k \in \bN_0},
 \end{align}
 for a monotone sequence of points in time \((t_k)_{k \in \bN_0}\) be given.
 We call \(\cT\) a \emph{Delaunay moving mesh}, iff \(\tau_k(t_k)\) is a Delaunay mesh for all \(t_k\), \(k \in \bN_0\).
\end{definition}
Obviously, if a moving mesh \(\cT\) is given with a Delaunay mesh \(\tau_k\), we cannot expect that \(\tau_k(t)\) is Delaunay for \(t > t_k\).
This happens if points move into the circumspheres of other cells and therefore are in conflict with them.
\begin{definition}[Conflict, conflict zone, resulting hole]
 Let a vertex set \(\cV\) and its corresponding Delaunay mesh \(\tau(\cV)\) be given.
 A point \(\widetilde{v} \in \bR^d\) is in \emph{conflict with a cell \(C \in \tau(\cV)\)} if it lays inside or on the boundary of the circumsphere of \(C\).
 We call the union of all cells that are in conflict with a point \(\widetilde{v}\) the \emph{conflict zone of  \(\widetilde{v}\)}.
 Furthermore, we call the union of all cells that are adjacent to a vertex \(v\in \cV\) the \emph{resulting hole of \(v\)}.
\end{definition}
Due to arising conflicts, a change of topology, meaning a redefinition of the cells, will be necessary to restore the Delaunay property.
This may change the number of cells.
Moreover, we even allow changing the number of vertices.
These topology changes take place at the discrete time points \((t_k)_{k \in \bN_0}\) from Definition~\ref{def:moving_mesh}.
The affected areas in this process will be the conflict zones and the resulting holes of the involved vertices.

We do not want to consider a moving mesh by itself.
Ultimately we are interested in performing numerical simulations with it.
More specifically, we focus on the class of finite volume methods. 
Therefore, we introduce the notion of generic, cell-wise defined data
\begin{align*}
 \UU(t) \coloneqq (\UU_1, \ldots, \UU_N) \in \bR^{N \times m},
\end{align*}
with \(N = \abs{\tau(t)} \in \bN\) being the number of all cells in \(\tau(t)\), and \(m \in \bN\) the dimension of the data.
Note that the data is implicitly related to the cell, i.e., the cell \(C_j\) corresponds to \(\UU_j \in \bR^{m}\), for all \(j = 1,\ldots, N\).
The data defines a function \(\UU \colon \Omega \times (0,\infty) \to \bR^{m}\) by
\begin{align}
\label{eq:data_function}
 \UU(x,t) \coloneqq \sum_{j = 0,\ldots,N} \UU_j \mathbf{1}_{C_j(t)}(x),
  \text{ with } C_j(t) \in \tau_k(t),
  t \in [t_k,t_{k+1}),  k \in \mathbb{N}_0,
\end{align}
where \(\mathbf{1}_{C_j}(x)\) denotes the indicator function of the cell \(C_j\).
\begin{remark}
Note that the notion of cell-wise constant data is only due to our focus on finite volume methods; other types of data are also possible.
In that case the treatment of the data in the following algorithms has to be adapted accordingly.
The connection to data can even be dropped altogether, without any impact on the presented algorithms.
\end{remark}
Due to the topology changes, we have to transfer the data from \(\tau_k(t_{k+1})\) to \(\tau_{k+1}(t_{k+1})\) in a meaningful way during the mesh adaption from \(\tau_k\) to \(\tau_{k+1}\).
Depending on the application the data can be either transferred in a conservative way via averaging or \(L^2\)-projections.
How this is done is explained in the following.

\begin{figure}[tp]
 \centering
 \subcaptionbox{\label{subfig:remeshing_0}}
 [0.45\linewidth]{
\includegraphics[width=0.65\linewidth]{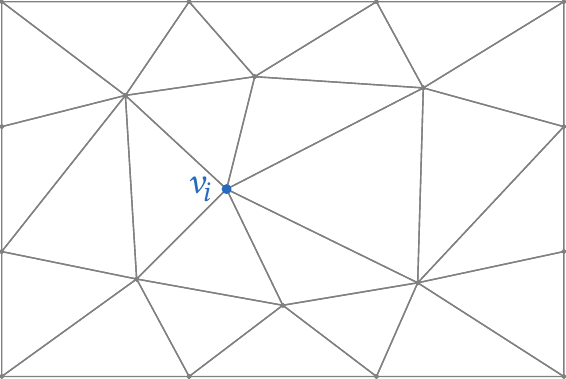}
 }\subcaptionbox{\label{subfig:remeshing_1}}
 [0.45\linewidth]{
\includegraphics[width=0.65\linewidth]{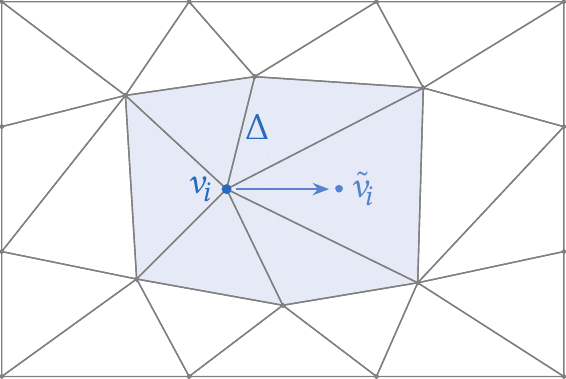}
 }

 \subcaptionbox{\label{subfig:remeshing_2}}
 [0.45\linewidth]{
\includegraphics[width=0.65\linewidth]{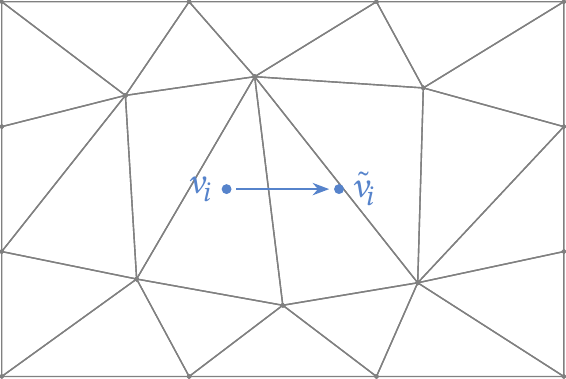}
 }\subcaptionbox{\label{subfig:remeshing_3}}
 [.45\linewidth]{
\includegraphics[width=0.65\linewidth]{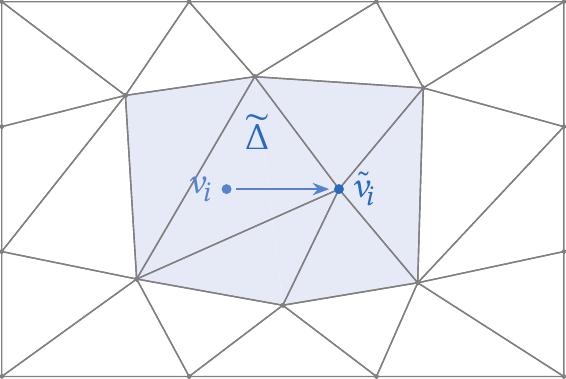}
 }\caption{Algorithm~\ref{alg:move}:
 (\subref{subfig:remeshing_0}) initial mesh,
 (\subref{subfig:remeshing_1}) get data from stencil \(\Delta\),
(\subref{subfig:remeshing_2}) remove vertex \(v_i\),
 (\subref{subfig:remeshing_3}) insert new vertex \(\widetilde{v}_i\), restore the Delaunay property, and computing the data on the new stencil \(\widetilde{\Delta}\).
 }
 \label{fig:move_vertex}
\end{figure}

For the presented mesh adaptions it suffices to consider the data \emph{locally} on a stencil \(\Delta \subset \tau\), which is a subset of connected cells \(C_j\) in a mesh \(\tau\).
During mesh adaptions, we encounter an old stencil \(\Delta\) (before the mesh adaption), and a new stencil \(\widetilde{\Delta}\) (after adapting the mesh) ---
see Figure~\ref{fig:move_vertex}  for (\subref{subfig:remeshing_1}) a sketch of an old stencil \(\Delta\), and (\subref{subfig:remeshing_3}) a sketch of a new stencil \(\widetilde{\Delta}\).
Both the old and the new stencil are used to compute the data on the resulting mesh.
We assume that both stencils are congruent in the sense that
\begin{align}
  \operatorname{dom}(\Delta) \coloneqq \bigcup_{C_j \in \Delta} C_j = \bigcup_{\widetilde{C}_j \in \widetilde{\Delta}} \widetilde{C}_j.
\end{align}
A standard choice for transferring the data \(\UU\) from an old stencil \(\Delta\) to a new stencil \(\widetilde{\Delta}\) on the new mesh would be the \(L^2\)-projection:
find \(\widetilde{\UU}\vert_{\operatorname{dom}(\Delta)}(x,t) = \sum_{\widetilde{C}_j \in \widetilde{\Delta}} \widetilde{\UU}_j \mathbf{1}_{\widetilde{C}_j}(x)\), such that
\begin{align}
  \label{eq:mmesh:l2projection}
 \int_{\operatorname{dom}(\Delta)} \norm[\big]{ \UU(x,t) - \widetilde{\UU}(x,t) }_2 \ddd x \longrightarrow \min,
\end{align}
where \(\UU(x,t)\) is defined as in \eqref{eq:data_function}.
Another way to transfer the data is by local averaging, i.e., setting the new data
\begin{align}
\label{eq:mmesh:local_avg_projection}
\widetilde{\UU}_j = \UU_{\Delta} \coloneqq
\frac{1}{\abs{\operatorname{dom}(\Delta)}}
    \sum_{C_j \in \Delta} \UU_j \abs{C_j},
\end{align}
for all \(\widetilde{C}_j \in \widetilde{\Delta}\).

From here on, we only consider Delaunay moving meshes, therefore, if we say \emph{moving mesh}, we always mean \emph{Delaunay moving mesh}.

In the following we present methods to manipulate vertices in a mesh (remove, insert, and move vertex), while retaining a Delaunay mesh and therefore a (Delaunay) moving mesh.
The vertex motion in context of Definition~\ref{def:moving_mesh} is hereby understood as an algorithmic transformation from \(\tau_k\) to \(\tau_{k+1}\), updating the vertex positions from time \(t_k\) to \(t_{k+1}\) in accordance to the vertex motions
\(m_v(t_{k+1})\).
In the following, we drop the time dependency, and focus on the motion of a single vertex \(v\) to its new position \(\widetilde{v} = v + m_v\) in a vertex set \(\cV_0 \subset \bR^d\).
\newline
The algorithmic implementation of the vertex-motion is based on two crucial functions:
first we \emph{remove} the vertex from the old position, then we \emph{insert} it at its new position.
It is important that both the insertion and the removal of the vertex consistently retain the cell-wise, underlying data.

The removal of a single vertex works as follows.
\begin{boxalgorithm}[label=alg:remove]{Remove Vertex}
  \textbf{Description:}
 Removes vertex \(v\in \cV\) in a Delaunay mesh \(\tau(\cV)\).
 \tcbsubtitle{Algorithm}
 \begin{enumerate}
  \item Old stencil \(\Delta\): Get data \(\UU\) and all cells \(C\) that are incident to \(v\).
        \label{item:old_stencil}
  \item Remove \(v\) from the Delaunay mesh \(\tau(\cV)\) and remesh the resulting hole, i.e., compute \(\tau\bigl(\cV\mathbin{\backslash}\{v\}\bigr)\).
        \label{item:modification}
  \item New stencil \(\widetilde{\Delta}\):
        Get all cells \(\widetilde{C}\) that are in conflict with the old vertex \(v\).
        \label{item:new_stencil}
  \item Compute data \(\widetilde{\UU}\) on the new stencil \(\widetilde{\Delta}\) given the data on the old stencil \(\Delta\).
 \end{enumerate}
\end{boxalgorithm}
Removing vertices from Delaunay meshes has been described in e.g. \cite{devillers.teillaud:perturbations:2003,devillers:vertex:2009}.
Removing a single vertex can be implemented with the computational complexity \(O(q \log q)\) \cite{devillers:deletion:2002}, where \(q \in \bN\) is the number of created cells.
In the worst case, the entire Delaunay mesh might be affected, leading to a complexity of \(O(N_{\cV}^{ \ceil{d/2} })\) with \(N_{\cV} = \abs{\cV}\), cf.  \cite{devillers:deletion:2002,clarkson.mehlhorn.ea:four:1993}.

The insertion of a point into an existing Delaunay mesh is similarly straightforward.
Again, we have to make sure that the data is transferred correctly during the mesh adaption.

\begin{boxalgorithm}[label=alg:insert]{Insert Vertex}
 \textbf{Description:}
 Insert vertex \(\widetilde{v} \in \Omega = \operatorname{conv}(\cV) \subset \bR^d\), in a Delaunay mesh \(\tau(\cV)\), with \(\widetilde{v} \notin \cV\).
 \tcbsubtitle{Algorithm}
 \begin{enumerate}
  \item Old stencil \(\Delta\): Get data \(\UU\) and all cells \(C \in \tau(\cV)\) that are in conflict with \(\widetilde{v}\).
  \item Insert \(\widetilde{v}\) into the Delaunay mesh.
        Remesh the conflict zone around \(\widetilde{v}\) and restore the Delaunay property.
  \item New stencil \(\widetilde{\Delta}\): Get all cells \(\widetilde{C}\) that are incident to \(\widetilde{v}\).
  \item Compute data \(\widetilde{\UU}\) on the new stencil \(\widetilde{\Delta}\) given the data on the old stencil \(\Delta\).
 \end{enumerate}
\end{boxalgorithm}

Inserting a vertex implies that a Delaunay mesh has to be computed, see e.g. \cite{boissonnat.devillers.ea:incremental:2009}.
Finding a Delaunay mesh for a given vertex set is equivalent to computing the convex hull for a lifted point set, cf. \cite{brown:voronoi:1979}.
As such, the complexity of inserting a vertex is at its worst  \(O(N_{\cV}^{ \ceil{d/2} })\) with \(N_{\cV} = \abs{\cV}\), see \cite{boissonnat.devillers.ea:incremental:2009}.
In practice such worst-case scenarios rarely occur, and if the mesh fulfills further (mild) requirements the expected runtime scales with \(O(N_{\cV} \log N_{\cV})\), as shown in \cite{clarkson.mehlhorn.ea:four:1993}.
\newline
To transfer the data on the new mesh, we use local averaging \eqref{eq:mmesh:local_avg_projection}, or alternatively \(L^2\)-projection \eqref{eq:mmesh:l2projection}.

Finally, moving a vertex is a combination of both methods above.
To preserve the locality of the algorithm and avoiding degenerating meshes, we restrict the motion distance of a vertex \(v \in \tau\) by the variable \(\omega(\tau, v) > 0\), which is given in such a way that the new position \(\widetilde{v}\) is guaranteed to be inside the polytope defined by the incident cells of \(v\), i.e.,
\begin{align}
  \label{eq:max_vertex_moving_distance}
 \omega(\tau, v) \coloneqq \min_{x \in \partial K} \operatorname{dist}(v,x),
 ~
 \text{ with }  K \coloneqq \bigcup_{\mathclap{
 \substack{C \in \tau, \\ C \text{ incident to } v }
 }} 
 \overline{C}.
\end{align}
For a given mesh \(\tau\) we can compute the overall minimum distance, denoted by
\begin{align}
  \label{eq:max_vertex_moving_distance:meshwise}
  \omega(\tau) \coloneqq \min_{v \in \cV} \omega(\tau, v).
\end{align}
Note that the maximum vertex motion distance is restricted by \(\omega(\tau)\) which in turn can be roughly estimated by the smallest insphere diameter of the mesh \(\tau\), i.e.,
\begin{align}
  \label{eq:mmesh:max_vertex_distance:insphere_estimate}
  \omega(\tau) \geq \min_{C \in \tau} (\operatorname{insphere\,diam}(C))
  \quad \text{ for all } v \in \cV,
\end{align}
where \(\operatorname{insphere\,diam}(C) \geq 0\) denotes the insphere diameter of a cell \(C\).
For nondegenerate meshes \eqref{eq:mmesh:max_vertex_distance:insphere_estimate} and if all insphere diameters can be uniformly bounded from below by a positive constant for all time, any mesh motion can be realized by  a finite number of time steps.

With this, we are able to formulate the algorithmic version of the vertex motion, which is also illustrated in Figure~\ref{fig:move_vertex}.

\begin{boxalgorithm}[label=alg:move]{Move Vertex}
\textbf{Description:}
 Move vertex \(v\in \cV\) to the new position \(\widetilde{v} \in \bR^d\) with
 \begin{align} \label{eq:move_restriction}
  \norm{v - \widetilde{v}}_2 < \omega(\tau, v)
 \end{align}
 in a Delaunay mesh \(\tau(\cV)\).
 \tcbsubtitle{Algorithm}
 \begin{enumerate}
  \item Old stencil \(\Delta\): Get data \(\UU\) and all cells \(C\) that are incident to \(v\) \emph{and} that are in conflict with \(\widetilde{v}\).
  \item Move vertex:
  \begin{enumerate}
    \item Remove \(v\) from the Delaunay mesh.
    \item Remesh the resulting hole.
    \item Insert \(\widetilde{v}\) into the Delaunay mesh.
    \item Remesh the conflict zone for \(\widetilde{v}\) and restore the Delaunay property.
  \end{enumerate}
  \item New stencil \(\widetilde{\Delta}\): Get all cells \(\widetilde{C}\) that are incident to \(\widetilde{v}\) \emph{and} that are in conflict with \(v\).
  \item Compute data \(\widetilde{\UU}\) on the new stencil \(\widetilde{\Delta}\) given the data on the old stencil \(\Delta\).
 \end{enumerate}
\end{boxalgorithm}

Algorithm~\ref{alg:move} is a combination of Algorithm~\ref{alg:remove}
and Algorithm~\ref{alg:insert}.
The data transfer involves no further cells despite those that are involved in the vertex removal\slash{}insertion.
As such, the additional computational costs are asymptotically negligible.
Consequently, the computational complexity of the algorithm is at most \(O\bigl(N_{\cV}^{ \ceil{d/2} }\bigr)\).

In practice \eqref{eq:move_restriction} is a time-step restriction, i.e., if a motion from \(v\) to \( \widetilde{v}\) does not satisfy \eqref{eq:move_restriction}, smaller time steps have to be chosen that satisfy \eqref{eq:move_restriction}.
Alternatively, other vertices that are too close to the new position could be removed (via Algorithm~\ref{alg:remove}) before moving the vertex.

Algorithm~\ref{alg:move} relies only on local operations within the stencil, which makes the algorithm highly efficient --- see Section~\ref{sec:mmesh_benchmark} for numerical benchmarks.
Furthermore, local mesh operations are the foundation for a possible parallelization of the whole method.

Algorithms~\ref{alg:remove}, \ref{alg:insert} and \ref{alg:move} are already sufficient for implementing \(r\)-adaptive mesh refinement (i.e., adaptivity by vertex relocation), albeit without considering interfaces within the mesh.
The steps~(\ref{item:old_stencil}), (\ref{item:modification}) and (\ref{item:new_stencil}) in all three algorithms can easily be implemented by using built-in methods of a Delaunay mesh implementation.

\begin{remark}
  \label{remark:mesh_quality}
 Generating well-shaped three- or \(d\)-dimensional meshes is no trivial task.
 In three dimensions there is the well-known phenomena of sliver tetrahedra (observed for example in \cite{cavendish.field.ea:apporach:1985}).
 These \emph{slivers} are almost degenerate tetrahedra whose vertices lie close to the equator of their circumsphere.
 We refer to \cite{cheng.dey.ea:sliver:2000,bern.chew.ea:dihedral:1995} for a more in-depth classification of badly shaped simplices.
 Even well-spaced vertices do not prevent the occurrence of slivers in Delaunay meshes \cite{talmor:well:1997}.
 \newline
 There are many methods to remove slivers and improve the mesh quality.
 For example, choosing appropriate weights in a weighted Delaunay mesh can remove slivers, see	\cite{cheng.dey.ea:sliver:2000}.
 This means no points are added\slash{}removed or perturbed, but the resulting mesh is not Delaunay anymore.
 Another approach is to perform Laplacian smoothing \cite{field:laplacian:1988} of the mesh, i.e., moving vertices to the midpoint of their neighbors.
 In this case vertices are moved, therefore the boundary and interface vertices have to be treated differently.
 Furthermore, Laplacian smoothing may degrade the mesh quality.
 This can be fixed by moving the vertices in order to  optimize some quality measures \cite{freitag.jones.ea:parallel:1999}.
 Slivers can also be removed by perturbing single vertices incident to slivers \cite{tournois.srinivasan.ea:perturbing:2009,edelsbrunner.li.ea:smoothing:2000}.
 The last approach in this list is to remove slivers by refining the mesh  \cite{li.teng:generating:2001,li:generating:2003}.
 The drawback hereby is that new points are added and insertions near mesh interfaces have to be treated separately.
 \newline
We note, that the occurrence of slivers in itself is not a problem when aiming at numerical simulations. 
For the case of finite volume methods sliver cells can be easily merged virtually with neighboring nondegenerate cells. 
\end{remark}

The algorithms in this section enable us to handle a moving mesh \(\cT\) algorithmically, given vertex motions \(m_v\) which respect the constraint
\eqref{eq:move_restriction}.
In Figure~\ref{fig:remeshing_steps} we illustrate the relevant steps.
During each time interval \(t \in [t_{k},t_{k+1})\) we define the
time-depending moving mesh \(\tau_k(t)\) according to the vertex motions
\(m_v(t)\) and Definition~\ref{def:moving_mesh}, retaining the topology of the mesh.
This yield the mesh \(\tau_k(t)\), for \(t \in [t_{k},t_{k+1})\), which may not be Delaunay anymore, or even close to degenerating.
This makes remeshing a necessity.
The transition from \(\tau_k\) to \(\tau_{k+1}\) --- including the data ---
is performed by applying Algorithm~\ref{alg:move} for every vertex in \(\tau_k\) corresponding to their motion \(m_v = m_v(t_{k+1})\).
Algorithmically this allows for topology changes and restores the Delaunay property of the resulting mesh \(\tau_{k+1}\).
\section{Interface Preserving Algorithms}
\label{sec:mmesh:interface_preserving}

The moving mesh methods (Algorithms~\ref{alg:remove}, \ref{alg:insert}, \ref{alg:move}) are the prerequisite for our main goal.
We consider the motion of an interface within a meshed domain.
The interface is represented as a subset of cell surfaces within the moving mesh.
Ideally, the remeshing process does not change the interface at all.
The succeeding algorithms ensure preservation of the interface topology such that after remeshing no two bulk vertices will be
connected by an edge crossing the new position of the interface.
To achieve interface preservation in this sense, and depending on the local properties of the prescribed motion, the position  of the interface at a some $\tau_k$ can slightly change.
The changes in the position of the interface lead to a geometrical error presumably of the order of the mesh size.
Altogether, 
this is what we call an \emph{interface preserving moving mesh}, which is our main novel contribution.
\begin{figure}[tp]
 \centering
\includegraphics[width=0.33\linewidth]{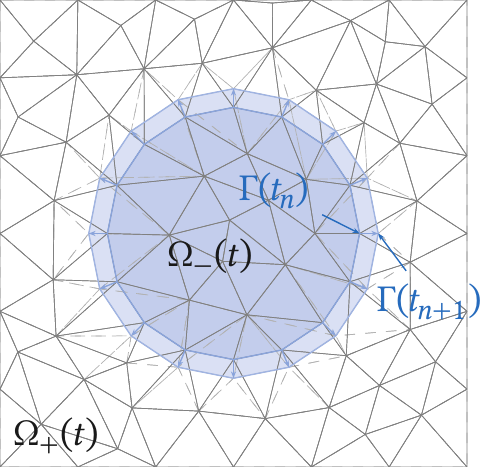}
 \caption{Sketch of a moving interface \(\Gamma(t)\) that divides two time-dependent subdomains \(\Omega_{\phaser}(t)\) and \(\Omega_{\phasel}(t)\).}
 \label{fig:two_phase_moving_mesh}
\end{figure}

First, we define an interface within a mesh.
\begin{definition}[Mesh Interface]
 \label{def:static_interface}
 An \emph{interface \(\Gamma_\tau\) in a mesh} \(\tau\) is a 
 \((d-1)\)-dimensional mesh  
 that consists of \((d-1)\)-dimensional facets of the cells in \(\tau\), which form a surface in \(\mathbb{R}^d\).
 It is denoted by \(\Gamma_\tau\). 
 The set of all vertices \(v \in \Gamma_{\tau}\) is denoted by \(\cV_\Gamma\).
\end{definition}
Note that the upper definition excludes the case of an interface with junction points.
Again, in principle, the algorithm may be extended to such cases, but requires special treatment of junctions.
However, we refrain from doing so, as the focus of this work is on closed interfaces for the description of phase transitions.

Definition~\ref{def:static_interface} addresses the static case and will serve as an initial state for the interface preserving moving mesh.
The construction of an initial (and thus static) surface interfaces within a full dimensional surrounding mesh may be realized in various well-known ways, e.g. 
by level-set approaches or direct parametrization.

In the next step we define a moving interface.
\begin{definition}[Moving Mesh Interface]
  \label{def:moving_interface}
 Let a moving mesh
 \begin{align}
   \cT  =  (\tau_k, \{m_v\}_{v\in \cV_k})_{k \in \bN_0}
 \end{align}
 and an (initial) interface \(\Gamma_{\tau_{0}}\) be given.
 We call 
 \begin{align}
  \Gamma_{\cT} 
  \coloneqq \bigl(
  \Gamma_{\tau_{k}},\{m_v\}_{v\in  \cV_{k} \cap \cV_{\Gamma_{\tau_{k}}} }
  \bigr)_{k \in \bN_0}
 \end{align}
 a \emph{moving mesh interface} if the following holds true.
\begin{itemize}
\item The vertices of the interface \(\Gamma_{\tau_{k}}\) move in time according to the corresponding vertices of the  moving mesh \(\tau_k(t)\). 
This defines the time-dependent interface \(\Gamma_{\tau_{k}}(t)\), \(t \in [t_k, t_{k+1})\).
\item The interface \(\Gamma_{\tau_{k}}(t)\) as a polygonal shape is retained during each transition from \(\tau_k(t_{k+1})\) to \(\tau_{k+1}(t_{k+1})\), for each \(k \in \bN_0\).
That means that \(\Gamma_{\tau_{k}}(t_{k+1})\) equals \(\Gamma_{\tau_{k+1}}(t_{k+1})\) as sets in \(\mathbb{R}^d\).
\end{itemize}
\end{definition}
A direct consequence of Definition~\ref{def:moving_interface} in conjunction with Definition~\ref{def:moving_mesh}, is that the interface retains its topology globally in time.
In particular, if the interface is closed the encapsulated  domain \(\Omega_\phasel(t)\) (cf. Figure~\ref{fig:two_phase_moving_mesh}) stays connected for all time.

As stated in Section~\ref{sec:moving_mesh}, to restore the Delaunay property of the entire mesh, we have to adapt the mesh topology. 
Notably, this should be done in a fashion, such that the moving mesh interface topology remains invariant.
Strictly speaking, our moving mesh method is not a moving interface in the sense of Definition~\ref{def:moving_interface}, instead we use the following relaxed definition of a moving interface.

\begin{definition}[\(\varepsilon\)-Moving Mesh Interface] 
  \label{def:eps_moving_interface}
 Let \(\varepsilon \geq 0\), a moving mesh
 \begin{align}
   \cT  =  (\tau_k, \{m_v\}_{v\in \cV_k})_{k \in \bN_0},
 \end{align}
 and an (initial) interface \(\Gamma_{\tau_{0}}\) be given.
 We call 
 \begin{align}
  \Gamma_{\cT} 
  \coloneqq \bigl(
  \Gamma_{\tau_{k}},\{m_v\}_{v\in  \cV_{k} \cap \cV_{\Gamma_{\tau_{k}}} }
  \bigr)_{k \in \bN_0}
 \end{align}
 an \emph{\(\varepsilon\)-moving mesh interface} if the following holds true.
\begin{itemize}
\item The vertices of the interface \(\Gamma_{\tau_{k}}\) move in time according to the corresponding vertices of the  moving mesh \(\tau_k(t)\). 
\item The interface \(\Gamma_{\tau}\) is a single  closed surface.
  \item \textbf{Approximation property:} For each \(k \in \bN_0\) the \(\varepsilon\)-neighborhood in \(\mathbb{R}^d\) of the interface of \(\tau_{k+1}(t_{k+1})\) contains the interface of \(\tau_k(t_{k+1})\).
  \item \textbf{Preservation property:} No two non-interface vertices (also called \emph{bulk} vertices) are connected by an edge which intersects the interface of \(\tau_{k+1}(t_{k+1})\).
\end{itemize}  
We will succeed to construct an interface motion algorithm by exploiting the stronger Gabriel property \cite{gabriel.sokal:new:1969} for interface facets.
All subsequent algorithms will result in a moving mesh with an $\varepsilon$-moving mesh interface.
\begin{definition}[Gabriel]
 Let a \((d-1)\)-dimensional facet \(\facet\) in a mesh be given.
 We call a sphere a \emph{witness sphere} of \(\facet\) if it contains all vertices of \(\facet\) on its boundary and does not contain any vertex of the mesh in its interior.
A facet has the \emph{Gabriel property} if its smallest circumscribing sphere is a witness sphere.
The smallest circumscribing sphere of a facet is called \emph{Gabriel sphere}.
\end{definition}
We note that any facet that has the Gabriel property is part of the Delaunay mesh.

\subsection{Interface Motion}
\label{sec:move_interface}

We propose an algorithm, where only interface vertices  are moved, which is based on Algorithm~\ref{alg:move}.
As for the vertex motion, the algorithmic treatment of the interface motion is understood in the context of the remeshing from the base mesh \(\tau_k\) to \(\tau_{k+1}\) (cf. Definition~\ref{def:moving_mesh}).

The main idea of the interface motion algorithm is depicted in Figure~\ref{fig:move_interface} and reads as follows.
Before moving interface vertices, we detect all non-interface vertices in the Gabriel spheres of all interface facets at their new positions and delete all of these non-interface vertices.
If additionally there are no interface vertices inside the Gabriel sphere then the interface facets must be Gabriel after interface motion.
As mentioned before, the interface facets are consequently also part of the Delaunay mesh after the vertex motion.
Nevertheless, if there are further interface vertices inside the Gabriel sphere then a remeshing of the interface surface is possible.
However, the preservation property of the interface is guaranteed which we will later on prove within Theorem~\ref{thm:minsphere} and Corollary~\ref{cor:minsphere}.
The resulting interface is an $\varepsilon$-moving mesh interface.
Therefore, even though the moving interface facets may be affected by topology changes resulting from the reconstruction of the Delaunay property the interface surface is preserved according to Definition~\ref{def:eps_moving_interface}.
\begin{figure}[tp]
 \centering
 \subcaptionbox{\label{subfig:interface_remeshing_0}}
 [0.45\linewidth]{
\includegraphics[width=0.65\linewidth]{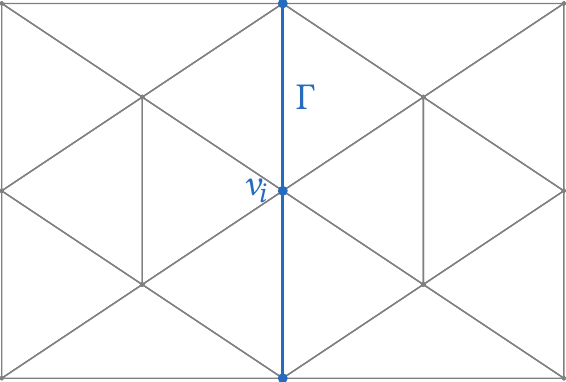}
 }\subcaptionbox{\label{subfig:interface_remeshing_1}}
 [0.45\linewidth]{
\includegraphics[width=0.65\linewidth]{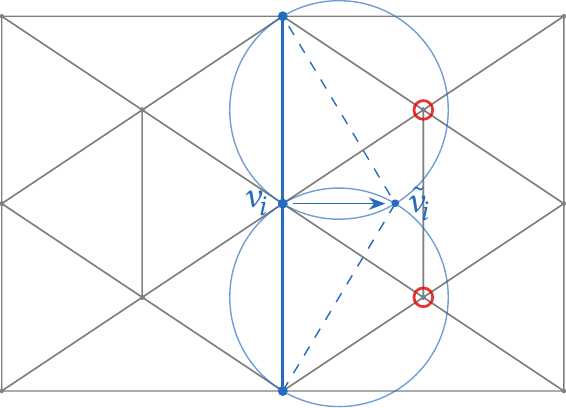}
 }

 \subcaptionbox{\label{subfig:interface_remeshing_2}}
 [0.45\linewidth]{
\includegraphics[width=0.65\linewidth]{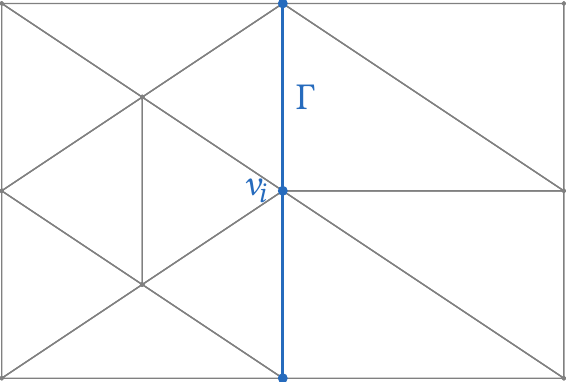}
 }\subcaptionbox{\label{subfig:interface_remeshing_3}}
 [.45\linewidth]{
\includegraphics[width=0.65\linewidth]{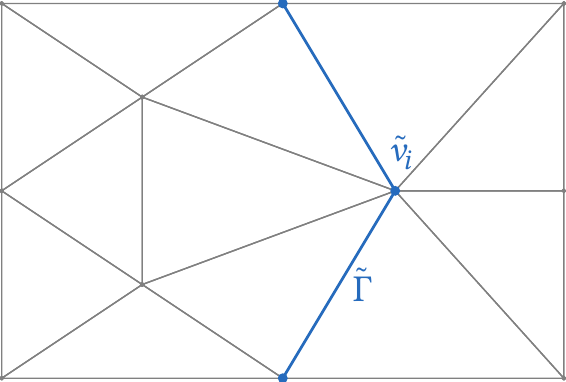}
 }\caption{Algorithm~\ref{alg:move_interface}:
 (\subref{subfig:interface_remeshing_0}) initial state,
 (\subref{subfig:interface_remeshing_1}) detecting non-interface vertices in conflict with Gabriel spheres,
 (\subref{subfig:interface_remeshing_2}) state after removing,
 (\subref{subfig:interface_remeshing_3}) state after moving the interface vertex.
 }
 \label{fig:move_interface}
\end{figure}

The Gabriel property alone, however, does not prevent the collision of vertices.
Consequently, we have to ensure that two vertices are always at least \(\xmin > 0\) apart from each other.
To achieve this, we traverse all interface vertices \(v \in \cV_\Gamma(t)\), and compute the distances to each adjacent vertex \(\widehat{v}\) of \(v\).
If an adjacent vertex \(\widehat{v}\) is closer than \(\xmin\) to \(v\) and does not belong to the interface, i.e., \(\widehat{v} \notin \cV_\Gamma(t)\), we remove it using Algorithm~\ref{alg:remove}.
For the coarsening of interface vertices we refer to Section~\ref{sec:interface_coarsening}.
\newline
So far, vertices are only moved or removed.
As a consequence the mesh only turns coarser in time.
A simple solution of this drawback is to store the removed non-interface vertices in a list \(L_{\mathrm{del}}\) and insert them whenever they are not inside a Gabriel sphere of any interface facet and are at least \(\xmin\) apart from any interface vertex.
In that way, we keep track of the initial mesh vertices.
This means that the initial mesh can be understood as a background mesh.

\begin{boxalgorithm}[label=alg:move_interface]{Move Interface}
\textbf{Description:}
 Move all interface vertices \(v \in \cV_\Gamma(t)\) to their new positions \(\widetilde{v} \in \bR^d\), with \(\norm{v - \widetilde{v}}_2 <  \omega(\tau) / 2\) (as defined in \eqref{eq:max_vertex_moving_distance:meshwise}), in a Delaunay mesh.

 \textbf{Parameters:}
 Minimum interface distance \(\xmin > 0\).

 \tcbsubtitle{Algorithm}
 \begin{itemize}
    \item \textbf{Ensure interface:} For all \(v \in \cV_\Gamma(t)\), do: Find all non-interface vertices that are inside the Gabriel spheres of all interface facets that are incident to \(v\) with new interface vertex position \(\widetilde{v} \in \bR^d\).
    Remove (Algorithm~\ref{alg:remove}) all detected non-interface vertices,
    while appending the position of the vertices to the list \(L_{\mathrm{del}}\).
    If the new vertex \(\widetilde{v} \in \bR^d\) is outside the domain \(\Omega\), throw an exception (\emph{boundary conflict}).
  \item \textbf{Move:} For all \(v \in \cV_\Gamma(t)\), do: Move (Algorithm~\ref{alg:move}) the interface vertex \(v\) to \(\widetilde{v}\).
  \item \textbf{Coarse}: Remove (Algorithm~\ref{alg:remove}) all non-interface vertices \(\widehat{v} \in \cV(t) \mathbin{\backslash} \cV_\Gamma(t)\) that are adjacent to an interface vertex \(v \in \cV_\Gamma(t)\) and fulfill   
  \( \norm{v - \widehat{v}}_2 < \xmin\).
  Append all removed vertices to the list \(L_{\mathrm{del}}\).
  \item \textbf{Refine}: Insert (Algorithm~\ref{alg:insert}) all vertices from the list \(L_{\mathrm{del}}\), that are not inside a Gabriel sphere of any interface facet and are at least \(\xmin\) apart from any interface node.
 \end{itemize}
\end{boxalgorithm}

In comparison to the Algorithm~\ref{alg:move} (move single vertex) we now move a whole set of vertices with Algorithm~\ref{alg:move_interface}.
Therefore, we took a stronger restriction on the vertex displacement.
Precisely speaking we check in Algorithm~\ref{alg:move_interface} that the vertices are not moved further than \(\omega(\tau) / 2 \) which is sufficient to ensure that no vertices collide,
or move out of their incident cell polytope \(K\), see \eqref{eq:max_vertex_moving_distance}.

The exception (\emph{boundary conflict}) points to an interaction of the domain boundary with the interface and has to be handled by a separate {(sub-)} model.

In Algorithm~\ref{alg:move_interface} no explicit remeshing is mentioned.
However, every call of the remove, insert and move Algorithms contains a local remeshing step.
Executing all those remeshing operations guarantees that the resulting mesh is still Delaunay after moving the interface.

As mentioned before Algorithm~\ref{alg:move_interface} gives a moving mesh with an \(\varepsilon\)-moving mesh interface.
In other words the interface is not exactly retained as polygonal shape.
Instead, the interface \(\tau_{k+1}(t_{k+1})\) resulting from moving and remeshing contains the interface \(\tau_{k}(t_{k+1})\) in its \(\varepsilon\)-neighborhood, which is the interface resulting from the movement alone.
The exact value of \(\varepsilon\) is not an input parameter but a result of the corresponding motion and is limited by the mesh width.
Only in special cases in two space dimensions (interface is not close to any intersection conflict) the interface remains invariant under the more strict Definition~\ref{def:moving_interface}, see also Remark~\ref{rem:minsphere} for a detailed explanation.

Since we keep track of deleted vertices in \(L_{\mathrm{del}}\) the initial mesh returns (locally) to its initial configuration as soon as the interface has passed through and is sufficiently far away from the considered region.
More precisely, a cell corresponds to a cell from the initial mesh, if its circumsphere does not contain any deleted (background) vertex from \(L_{\mathrm{del}}\).
In the two-dimensional case, those are all cells with circumsphere which have a distance of at least
\begin{align}
\label{eq:dist_background_mesh}
\max\Bigl( \xmin, ~ \max_{\substack{\text{interface} \\ \text{facet } S}}\bigl(\operatorname{length}(S)\bigr) \Bigr)
\end{align}
to the interface.

Using Algorithm~\ref{alg:move_interface}, we are now able to define an interface \(\Gamma_{\cT}(t)\) preserving moving mesh \(\cT\).
The steps involved are illustrated in Figure~\ref{fig:interface_remeshing_steps}.
Starting with a base-mesh \(\tau_k\), we have a time-dependent sub-mesh \(\tau_k(t)\) with interface \(\Gamma_{\tau_k}(t)\) on the time interval \(t \in [t_k, t_{k+1})\).
At the end of this time interval, at \(t_{k+1}\), we generate the new base-mesh \(\tau_{k+1}\) from \(\tau_k\) and \(\Gamma_{\tau_k}(t_{k+1})\), by means of Algorithm~\ref{alg:move_interface}.

\begin{figure}[tp]
 \centering
\includegraphics[width=0.75\linewidth]{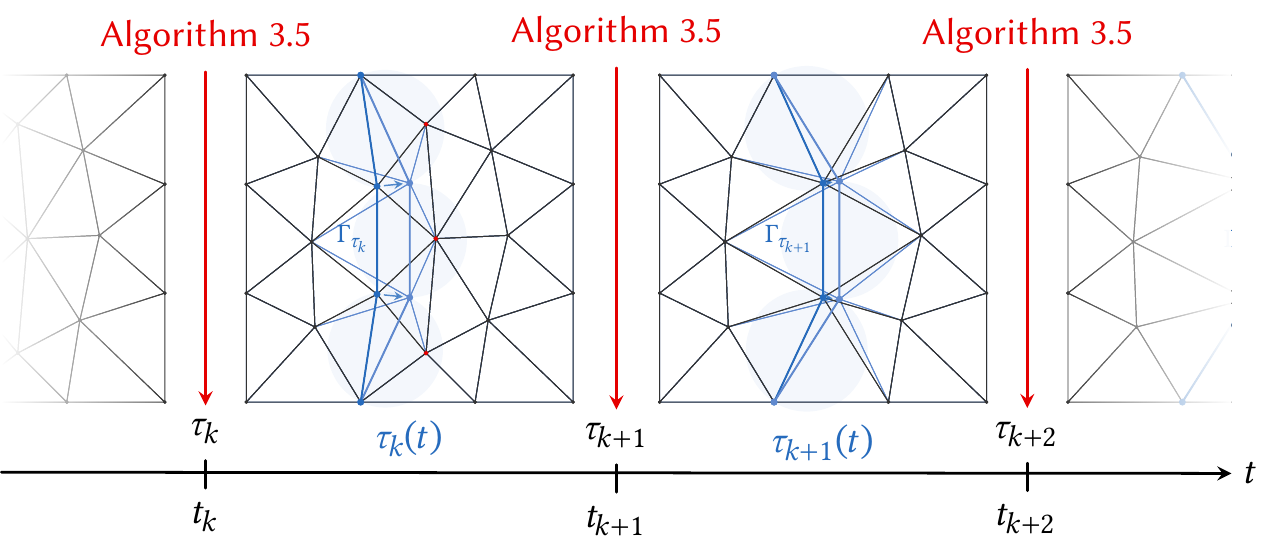}
 \caption{Illustration of an interface preserving moving mesh in time.}
 \label{fig:interface_remeshing_steps}
\end{figure}

\subsection{Interface Refinement and Coarsening}
\label{sec:interface_refine_coarse}

During the interface motion, very large or very small interface facets can occur.
This can have a negative impact on numerical simulations with moving interfaces.
To retain a high-quality interface mesh, it is important to be able to refine and coarsen the interface mesh.
\end{definition}

We present two algorithms for interface mesh refinement and coarsening in Algorithms~\ref{alg:refine_interface} and \ref{alg:coarsen_interface} respectively.
We note that these algorithms are able to run in conjunction with an adaptive mesh refinement\slash{}coarsening in the bulk mesh.

\subsubsection{Interface Refinement}
\label{sec:interface_refinement}

The interface \(\Gamma_\tau\) is refined by refining its facets.
To decide which facets should be refined, we introduce the threshold \(\xmaxInterface > 0\).
If a facet has an edge that is larger than \(\xmaxInterface\) it is refined by inserting the midpoint of the longest edge as an additional interface vertex.

\begin{boxalgorithm}[label=alg:refine_interface]{Refine Interface}
 \textbf{Description:}
 Refine all interface facets that have an edge larger than the threshold value \(\xmaxInterface > 0\), while retaining the mesh interface.

 \textbf{Parameters:}
  edge length threshold \(\xmaxInterface > 0\).
 \tcbsubtitle{Algorithm}
For all facets \(S\) in \(\Gamma_{\tau}\) do
\begin{itemize}
  \item Compute the maximum  edge length \(l_{\max}\) of all edges \(e\) of \(S\). 
\item If \(l_{\max} > \xmaxInterface\): Insert (Algorithm~\ref{alg:insert}) the midpoint of the longest edge of \(S\) into the mesh.
\end{itemize}
\end{boxalgorithm}

The insertion of a new interface vertex usually result in remeshing. 
Nevertheless, we do not need to ensure the Gabriel property of the new interface facets,
as opposed to the interface motion Algorithm~\ref{alg:move},  due to the following reason:
\newline
Before inserting the new interface facet there were no two bulk vertices from different bulk domains that were connected via an edge. 
This means that there is no witness sphere for any pair of such vertices. 
After the insertion of an additional interface vertex there are even more restrictions on the witness sphere and therefore the connecting edge cannot exist after the insertion either. 
This means that the interface surface is preserved according to Definition~\ref{def:eps_moving_interface}.
In the two-dimensional setting the interface is even invariant according to the stronger Definition~\ref{def:moving_interface}.

\subsubsection{Interface Coarsening}
\label{sec:interface_coarsening}

To coarsen the interface, we remove single interface vertices from the mesh. To decide which facet should be coarsened we introduce another threshold \(\xminInterface > 0\).
As for the interface refinement, we decide whether an interface vertex should be removed, if an interface edge is shorter than \(\xminInterface\).
In this case we remove one of its vertices.
Before removing the interface vertex, we require some mesh modifications to preserve the interface which is guaranteed via the minimum covering sphere.

\begin{definition}[Minimum covering sphere]
 Let a non-empty set of vertices \(\cV = \{v_1, \ldots, v_m\}\) be given.
 We call a sphere a \emph{minimum covering sphere} of \(\cV\) if it is the smallest sphere which contains all vertices of \(\cV\) inside or on its boundary.
\end{definition}

\begin{remark}
The minimum covering sphere is unique and can be computed in linear time $\mathcal{O}(m)$ \cite{megiddo:linear:1983}. \newline
Assume that $\{v_1, \ldots, v_d\}$ are the $d$ vertices of a non-degenerate facet in a $d$-dimensional mesh.
Then the minimum covering sphere contains all vertices at its boundary and coincides with the Gabriel sphere.
\end{remark}

\begin{figure}[tp]
\centering
\includegraphics[width=0.70\linewidth]{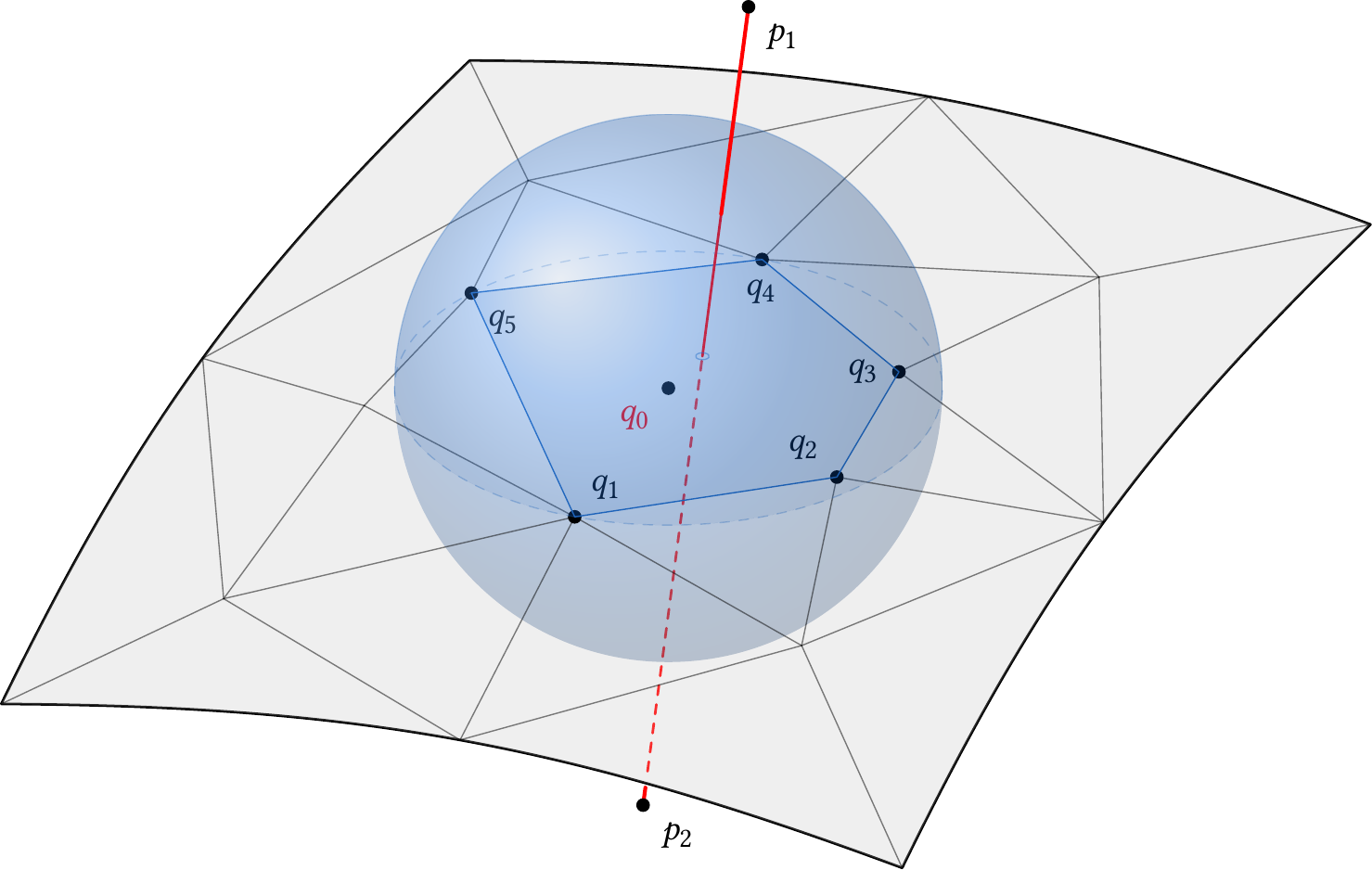}
\caption{Minimum covering sphere of the vertex set \(\{v_1, \ldots, v_5\}\). The set \(\{v_1, \ldots, v_5\}\) is hereby the resulting hole in a two-dimensional interface after the removal of the interface vertex $v_0$.}
\label{fig:sketch_minsphere}
\end{figure}

Assume that an interface vertex $v_0$ with adjacent interface vertices $\{v_1, \ldots, v_m\}, m \geq d$ as in Figure~\ref{fig:sketch_minsphere} has to be deleted.
Then there is a resulting hole in the interface mesh that will be remeshed automatically by the Delaunay mesh.
A priori, we do not know how the resulting hole will be remeshed.
However, we need to preserve the interface surface -- in other words we need to ensure that no two bulk vertices from two different phases will be connected via an edge running through the new resulting hole.
We do this by the following simple approach:
\begin{itemize}
 \item Build the minimum covering sphere containing the resulting hole, i.e., all vertices $v_1, \ldots, v_m$.
 \item Remove all bulk vertices that are inside or on the minimum covering sphere.
\end{itemize}
The interface surface is then preserved. We first state the resulting Algorithm~\ref{alg:coarsen_interface} and subsequently prove the preservation property within the following Theorem~\ref{thm:minsphere} and Corollary~\ref{cor:minsphere}.

\begin{boxalgorithm}[label=alg:coarsen_interface]{Coarsen Interface}
 \textbf{Description:}
 Coarsen all interface facets that have an edge smaller than the threshold value \(\xminInterface > 0\), while retaining the mesh interface.

 \textbf{Parameters:}
  edge length threshold \(\xminInterface > 0\).
\tcbsubtitle{Algorithm}
For all vertices \(v\) in \(\Gamma_{\tau}\) do
\begin{itemize}
\item Compute the minimum edge length \(l_{\min}\) and average edge length \(l_{\mathrm{avg}}\) of all interface edges \(e\) incident to \(v\). 
\item If \(l_{\min} < \xminInterface\): Write the vertex \(v\) and the average edge length \(l_{\mathrm{avg}}\) into a sorted list \(L_{\mathrm{coarsen}}\) with ascending \(l_{\mathrm{avg}}\).
\end{itemize}
While \(L_{\mathrm{coarsen}}\) is not empty do
\begin{itemize}
    \item Take the first vertex \(v\) of \(L_{\mathrm{coarsen}}\) and remove it from the list.
    \item Build the minimum covering sphere containing the resulting hole of \(v\) in \(\Gamma_{\tau}\), i.e., all incident interface vertices.
    \item Remove (Algorithm~\ref{alg:remove}) all bulk vertices from the mesh that are inside and on the minimum covering sphere.
    \item Remove (Algorithm~\ref{alg:remove}) the interface vertex \(v\) from the mesh.
    \item Update the list \(L_{\mathrm{coarsen}}\):
    For all interface vertices \(\widehat{v}\) adjacent to \(v\), compute the new minimum edge length \(l_{\min}\) and the new average edge length \(l_{\mathrm{avg}}\).
    If \(l_{\min} \geq  \xminInterface\), delete the vertex \(\widehat{v}\) from the list.
    Otherwise, update the position of the vertex \(\widehat{v}\) in the list according to \(l_{\mathrm{avg}}\).
\end{itemize}
\end{boxalgorithm}

\begin{remark}
\label{rem:minsphere}
Algorithm~\ref{alg:coarsen_interface} works in arbitrary space dimensions and generalizes the following procedure in a two-dimensional mesh:
In two space dimensions we always get as resulting hole a segment $\overline{v_1 v_2}$ whose minimum covering circle is the Gabriel circle. 
Therefore, if all bulk vertices are removed from the Gabriel circle and additionally no other interface vertex lies inside the Gabriel circle, the segment $\overline{v_1 v_2}$ has the Gabriel property and is thus part of the mesh.
In this case the interface is obviously preserved.\newline
Both the refining and the coarsening algorithm use the edge length as an indicator whether the interface should be refined or coarsened, however, the algorithms are not restricted to this case and other indicators such as the local curvature can be employed.
\end{remark}

\begin{theorem}[Empty minimum covering sphere]
\label{thm:minsphere}
Let a set of vertices $\{v_1, \ldots, v_m\}$, $m \geq d$ and its minimum covering sphere $M$ be given.
Consider any two vertices $p_1$ and $p_2$ that are given outside $M$ with the segment $\overline{p_1 p_2}$ intersecting the convex hull of $v_1, \ldots, v_m$
(see Figure~\ref{fig:sketch_proof_minsphere} for the setting).

Then there is no witness sphere of the segment $\overline{p_1 p_2}$.
\end{theorem}

This leads directly to the desired property for interface preservation.

\begin{corollary}[Empty minimum sphere preserves interface] \label{cor:minsphere}
Let a Delaunay mesh \(\tau\) with vertex set \(\cV\) and interface \(\Gamma_\tau\) be given.
Let a subset \(Q \coloneqq \{v_1, \ldots, v_m\} \subset \cV_\Gamma\), \(m \geq d\), of interface vertices be given such that the minium covering sphere does not contain any other vertices than \(v_1, \ldots, v_m\).
For any other two vertices \(p_1, p_2 \in \cV \setminus Q\) the following statement holds.
If the segment \(\overline{p_1 p_2}\) intersects the convex hull of \(Q\), then the segment is not an edge in the Delaunay mesh \(\tau\).
\end{corollary}

The corollary is important for Algorithm~\ref{alg:coarsen_interface} in the following sense.
When removing a vertex from the interface we get a resulting hole in the interface mesh.
We take the interface vertices of the resulting hole as the set \(Q\) in the Corollary~\ref{cor:minsphere} and compute their minimum covering sphere.
Corollary~\ref{cor:minsphere} ensures that no edge between two bulk vertices \(p_1\) and \(p_2\) crosses the newly meshed interface by removing all bulk vertices that are inside the minimum covering sphere.

\begin{proof}
We give the proof of Theorem~\ref{thm:minsphere} for $d=3$. 
For the other cases, we refer to the remark below.
The proof is carried out by contradiction.
Assume that $\overline{p_1 p_2}$ has a witness sphere $W$.
This implies that $W$ does not contain any of the points $v_1, \ldots, v_m$.
\begin{enumerate}
 \item First, we know that $\overline{p_1 p_2}$ has an intersection with $\operatorname{conv}(\{v_1,\ldots,v_m\})$. 
 Therefore, there is a facet of a triangulation of the convex hull, which intersects with $\overline{p_1 p_2}$ --- the special case that we hit an edge is discussed at the end of the proof.
 We call the vertices of this facet $v_a$, $v_b$ and $v_c$
 (in Figure~\ref{subfig:3d_coarsening_proof:1} this facet is for example given through $v_1$, $v_3$ and $v_5$).
We denote the intersecting point by $\ipnt$.
The intersecting point must then be both in the interior of the minimum covering sphere $M$ as well as in the interior of $W$.
\begin{figure}[tp]
\centering
\subcaptionbox{\label{subfig:3d_coarsening_proof:1} }
 [0.475\linewidth]{
\includegraphics[width=0.9\linewidth]{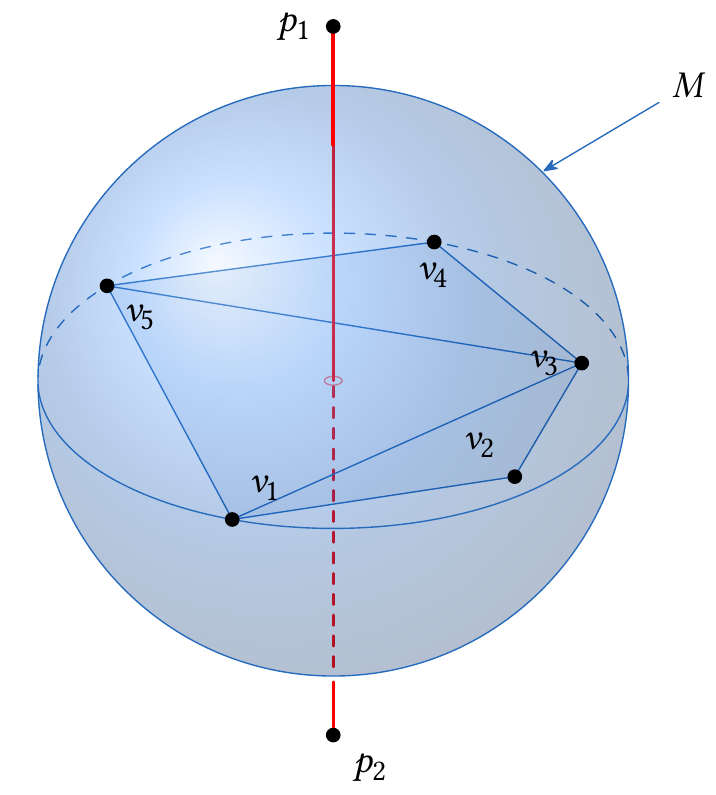}
}\subcaptionbox{\label{subfig:3d_coarsening_proof:2} }
[0.475\linewidth]{
\includegraphics[width=0.9\linewidth]{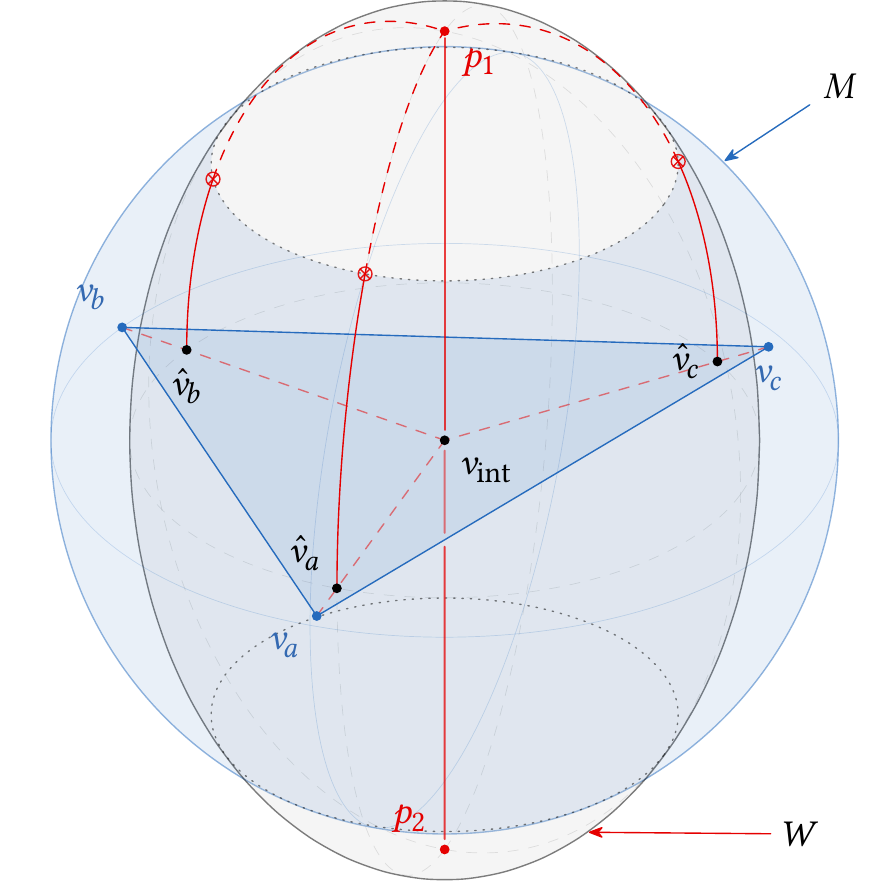}
}\caption{Exemplary sketch of (\subref{subfig:3d_coarsening_proof:1}) the setting and (\subref{subfig:3d_coarsening_proof:2}) the construction points of the proof of Theorem~\ref{thm:minsphere}.}
\label{fig:sketch_proof_minsphere}
\end{figure}
\item The points $v_a$, $v_b$ and $v_c$ are outside the sphere $W$, since it is a witness sphere. The segment from $\ipnt$ to $v_a$ goes from the inside of $W$ to the outside of $W$ and is fully located inside $M$.
Therefore, there is a point $\widehat{v}_a$ along the segment, which is on the sphere $W$ as well as inside the minimum covering sphere $M$. 
The same arguments hold for the vertices $v_b$ and $v_c$.
Thus, there are vertices \(\widehat{v}_a\), \(\widehat{v}_b\), \(\widehat{v}_c\) inside \(M\) and on \(W\).
\item For $p_1$ and $p_2$ we have by assumption on $M$ and by construction of $W$ that \(p_1\), \(p_2\) are outside \(M\) and on \(W\).
\item From (2) and (3) we know that $p_1$ and $\widehat{v}_a$ are on the sphere $W$, and thus we can find a geodesic on $W$ connecting both points. 
Furthermore, $p_1$ is outside $M$ and $\widehat{v}_a$ is inside $M$ and thus there must be an intersecting point of the geodesic with $M$. 
With this argument, we get six distinct points that are on both spheres $W$ and $M$, by relating $p_1$ with $\widehat{v}_a$, $\widehat{v}_b$, $\widehat{v}_c$ and $p_2$ with $\widehat{v}_a$, $\widehat{v}_b$, $\widehat{v}_c$.
If two spheres coincide on four or more points in general position they must be equal. 
This implies that $W = M$, which is a contradiction to $W$ being a witness sphere since $W=M$ contains all vertices $v_1, \ldots, v_m$.
\end{enumerate}
In the special case that in step (1) the intersecting point \(\ipnt\) lies on an edge and not a full facet, we get only two vertices \(v_a\) and \(v_b\) instead of three vertices.
However, we can use that the points \(p_1\), \(p_2\), \(v_a\), and \(v_b\) are coplanar.
Here, the witness sphere and the minimum covering sphere cannot be shown to be equal (like in step (4)), but their cross-section circles coincide in the plane and contain the vertices \(v_a\) and \(v_b\).
Therefore, even in this case the segment \(\overline{p_1 p_2}\) cannot have a witness sphere.
\end{proof}

One can see from the proof that both points $p_1$ and $p_2$ must be outside $M$. 
If only either $p_1$ or $p_2$ is outside $M$, we only get three points where the spheres coincide, which is not enough for concluding the equality.
 
\begin{remark}
The proof is independent of the spatial dimension.
The only changes are that the number of vertices of a facet changes to $d$ and the number of equal points of $\partial W$ and $\partial M$ changes to $2\cdot d$, which is obviously more than $d+1$.
If $d+1$ points in general position coincide for two $d$-dimensional spheres they must be equal.
\end{remark}

\section{Test Cases}
\label{sec:mmesh_benchmark}

For the implementation in two or three space dimensions most frameworks can be applied that are capable of inserting and removing points from a Delaunay mesh.
To implement the method we choose the CGAL-framework \cite{project:cgal:2020}, that provides these standard operations.
For background on the technical details and the implementation within CGAL we refer to  \cite{broennimann.fabri.ea:2d:2020,yvinec:2d:2020,pion.yvinec:2d:2020,jamin.pion.ea:3d:2020,rineau:2d:2020,alliez.jamin.ea:3d:2020,alliez.pion.ea:principal:2020}.
\newline
Our open source implementation of the moving mesh algorithm can be found at \cite{alkaemper:interface:2021}.

To test the algorithms, we consider several test cases in two and three space dimensions.
The algorithms \ref{alg:move}, \ref{alg:refine_interface} and \ref{alg:coarsen_interface} require the thresholds $\xmin$, $\xmaxInterface$ and $\xminInterface$.
In all test cases we choose
\begin{align*}
\xmin = 0.9\, h_{\mathrm{min}}, \quad
\xminInterface = (1 - p) h_{\Gamma,\mathrm{min}}, \quad
\xmaxInterface = (1 + p) h_{\Gamma,\mathrm{max}},
\end{align*}
with $h_{\mathrm{min}}$ the minimum edge length of all bulk edges and $h_{\Gamma,\mathrm{min}}$ and $h_{\Gamma,\mathrm{max}}$ the minimum and maximum edge length of all interface edges, respectively.
The percentage $p>0$ is chosen as smallest value such that the conditions
\begin{align*}
\xmaxInterface \geq 3\, \xminInterface \quad \text{and} \quad p \geq 0.1
\end{align*}
are fulfilled.
Note that for a sensible choice $\xmaxInterface$ must be chosen bigger than twice $\xminInterface$. 
Otherwise, the refinement of the interface would directly be undone in the interface coarsening step.
Furthermore, $p$ must be positive in order to guarantee that the interface is neither refined nor coarsened if no size changes occur.
To be robust in the presence of heterogeneous mesh sizes we compute $h_{\mathrm{min}}$, $h_{\Gamma,\mathrm{min}}$ and $h_{\Gamma,\mathrm{max}}$ as the 1\% and 99\% percentile of the edge lengths, respectively.

All computations in this section were performed sequentially on a desktop computer equipped with an AMD Ryzen Threadripper 2950X 16-core processor and \SI{128}{\giga\byte} RAM.
In the following we present the test cases.

\subsection{Star-Shaped Deformation of a Sphere} \label{subsec:mmesh_benchmark:star}

In this first test case, we start with a given interface \(\Gamma(0)\) and move its vertices according to a prescribed vector field 
\begin{align}
\label{eq:prescr_vec_field}
  M \colon [t_0, t_{\mathrm{end}}] \times \bR^d \to \bR^d : (t, \vx) \mapsto M(t,\vx), 
\end{align}
with the time interval from \(t_0 = 0\) to \(t_{\mathrm{end}} = 3\).
In each time step \(t_{k+1} = t_k + \Delta t\) with step size \(\Delta t > 0\), each interface vertex \(v \in \cV_{\Gamma}(t_k)\) is then moving according to 
\begin{align}
\label{eq:prescr_motion}
  v(t_{k+1}) = v(t_k) + \Delta t  M(v(t_k)).
\end{align}

We measure the performance of the interface moving Algorithm~\ref{alg:move_interface} in two and three space dimensions 
and compare it with the performance of the interface refinement and coarsening Algorithms~\ref{alg:refine_interface},
\ref{alg:coarsen_interface} as well as with the insertion and removal Algorithms~\ref{alg:remove},
\ref{alg:insert} for the bulk vertices.

\subsubsection{The Two-Dimensional Algorithm}
\label{subsec:mmesh_benchmark:2d:star}

\begin{figure}[tp]
 \centering
 \subcaptionbox{\label{subfig:2d:star:0} \(t=2\)}%
 [0.33\columnwidth]{%
 \includegraphics[width=0.33\columnwidth]{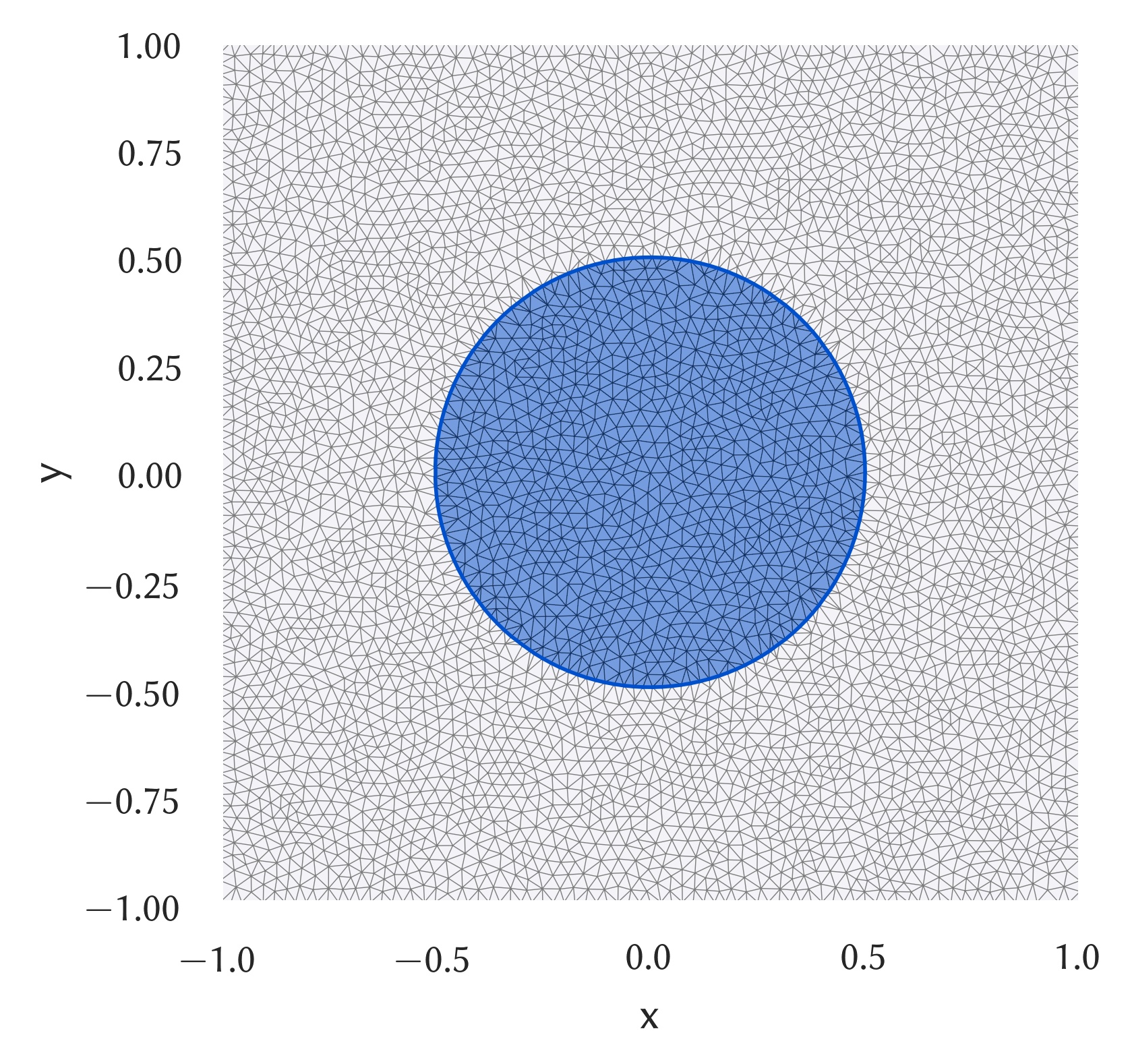}%
 }%
 \subcaptionbox{\label{subfig:2d:star:1} \(t=2.2\)}%
 [0.33\columnwidth]{%
 \includegraphics[width=0.33\columnwidth]{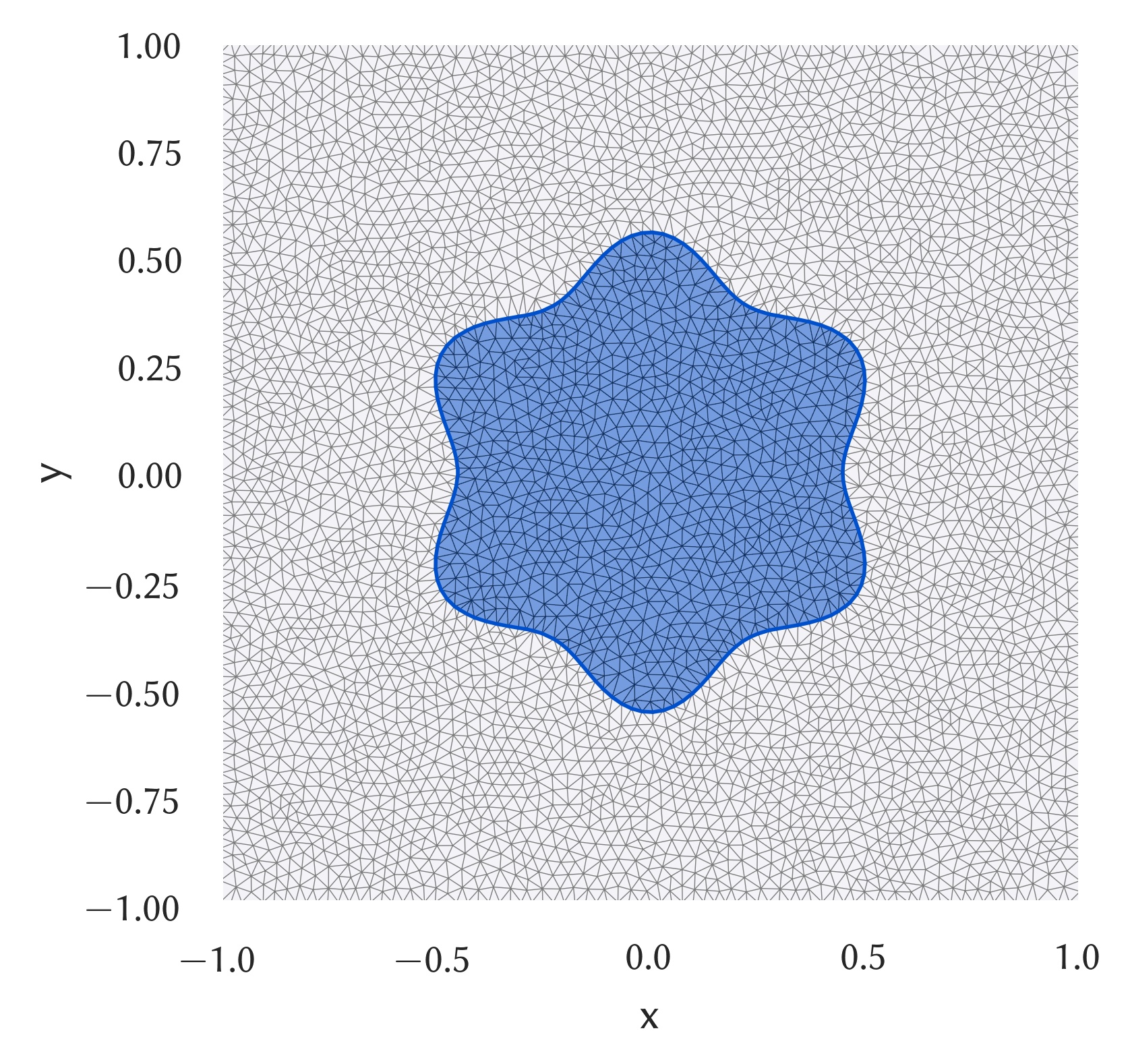}%
 }%
 \subcaptionbox{\label{subfig:2d:star:2} \(t=2.5\)}%
 [0.33\columnwidth]{%
 \includegraphics[width=0.33\columnwidth]{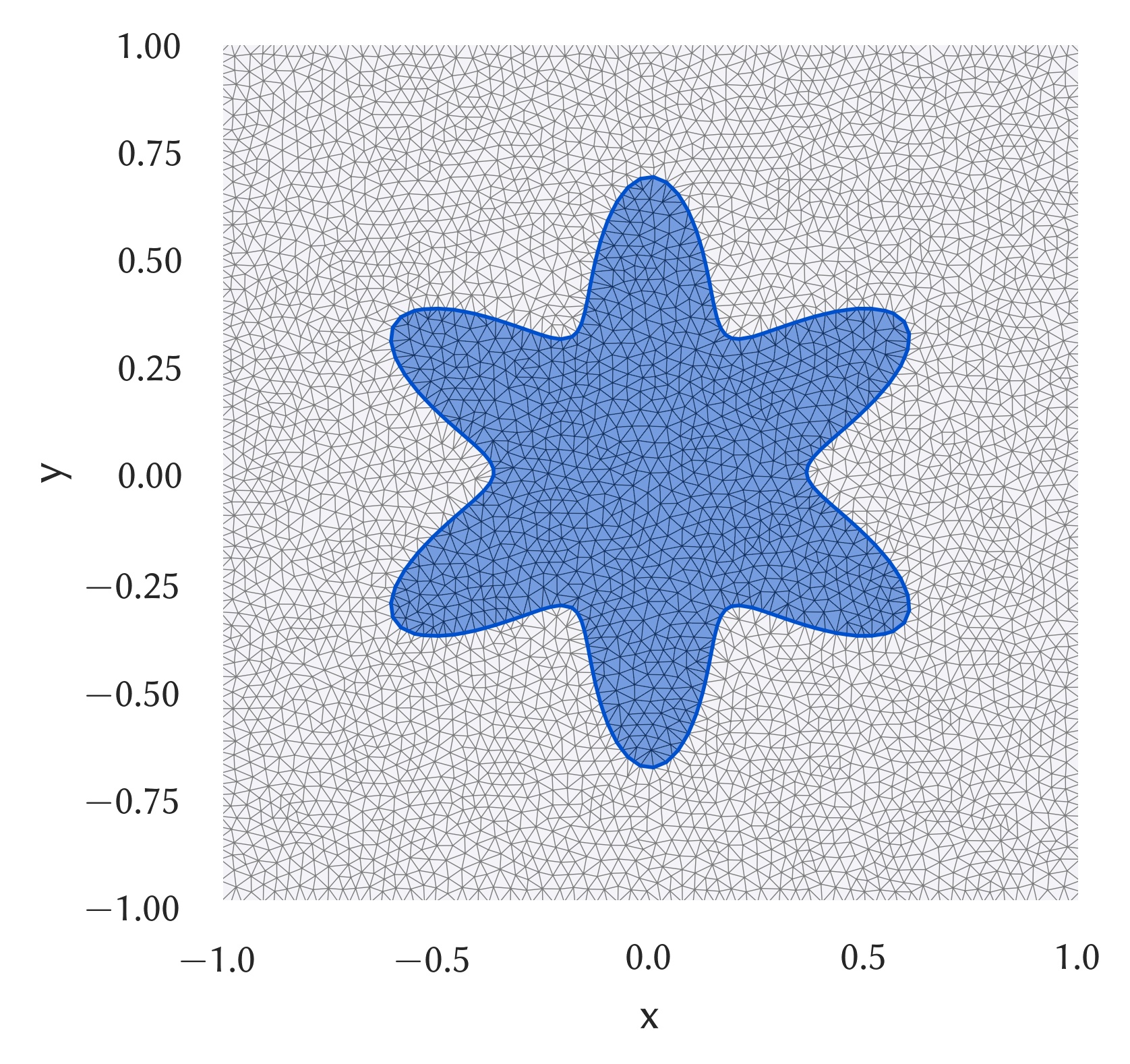}%
 }%
  \caption{%
    The moving interface (blue edges) for the star-shaped deformation example in two space dimensions.
    The depicted example was computed on a mesh with in average 8100 cells. }
 \label{fig:mmesh_benchmark:star:2d}
\end{figure}

\begin{figure}[tp]
 \centering
\includegraphics[height=0.33\linewidth]{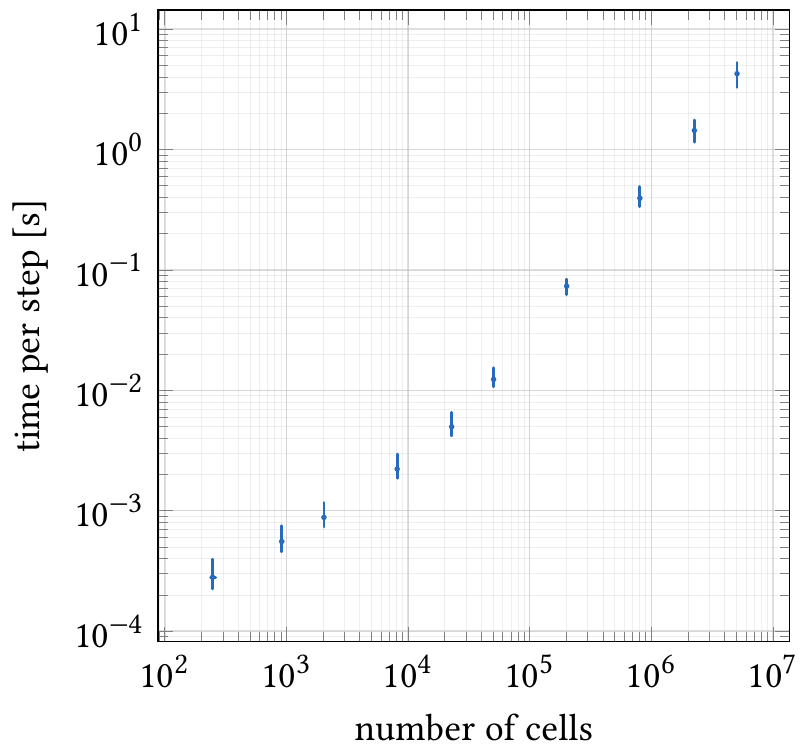}\hspace{0.1\columnwidth}\includegraphics[height=0.33\linewidth]{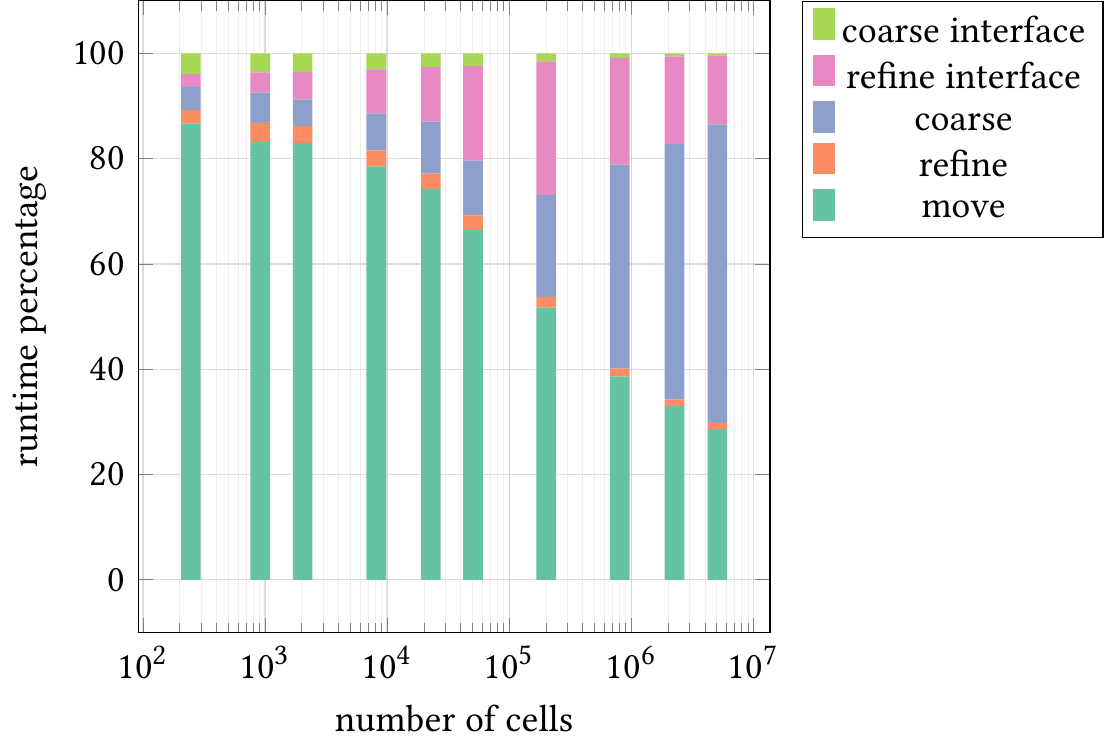}\caption{Computational time per step for the star-shaped deformation example in two space dimensions.
  In the left figure the lines represent the 90\%-percentile of the data over all time steps.
 The bars in the right figure indicate the mean computational time which is split into the respective mesh operation contributions:
move interface, refine/coarse (in Algorithm~\ref{alg:move_interface}), and refine/coarse interface.
 }
 \label{fig:2d:benchmark:star:runtime}
\end{figure}

For this test we consider a moving interface \(\Gamma(t)\) that starts as a circle with radius \num{0.5} and center in the origin.
The simulations run from \(t_0 = 0\) to \(t_{\mathrm{end}} = 3\) with time steps \(\Delta t = \num{2e-4}\). 
In each time step all interface points are moved according to the vector field
\begin{align}
  M \colon [t_0, t_{\mathrm{end}}] \times \bR^2 \to \bR^2 : (t, \vx) \mapsto -\sin(2 \pi t) \cos\bigl(2 \lceil t \rceil \operatorname{atan2}(x_2,x_1)\bigr) \vx.
\end{align}
This results in a deformation of the circle into a star-shaped curve and returns into circular shape for integer-valued times $t$, as seen in Figure~\ref{fig:mmesh_benchmark:star:2d}~(\subref{subfig:2d:star:0})--(\subref{subfig:2d:star:2}) for one simulation.
\newline
To test the algorithm, we generate meshes at several resolutions, parametrized by maximum edge lengths \(\Delta x > 0\) during the generation.
This example shows, as motivated in the introduction, that Algorithm~\ref{alg:move_interface} can handle strong interface deformations.
The interface length in Figure~\ref{fig:mmesh_benchmark:star:2d} varies in a range between \num{3.14} and \num{5.22} while the quality of the surrounding mesh does not differ from the quality of the initial one.
In fact, the surrounding mesh reverts to its initial configuration as soon as the distance to the interface is big enough, see \eqref{eq:dist_background_mesh}.
\newline
The computational time needed for each time step is shown in Figure~\ref{fig:2d:benchmark:star:runtime}.
From those figures one can see that the largest share of the runtime is asymptotically spent on the bulk coarsening.
Interface and bulk modifications, that allow for these strong deformations, are comparatively inexpensive.

\subsubsection{The Three-Dimensional Algorithm}
\label{subsec:mmesh_benchmark:3d:star}

\begin{figure}[tp]
 \centering
 \subcaptionbox{\label{subfig:3d:star:0} \(t=2\)}%
 [0.33\columnwidth]{%
 \includegraphics[width=0.33\columnwidth]{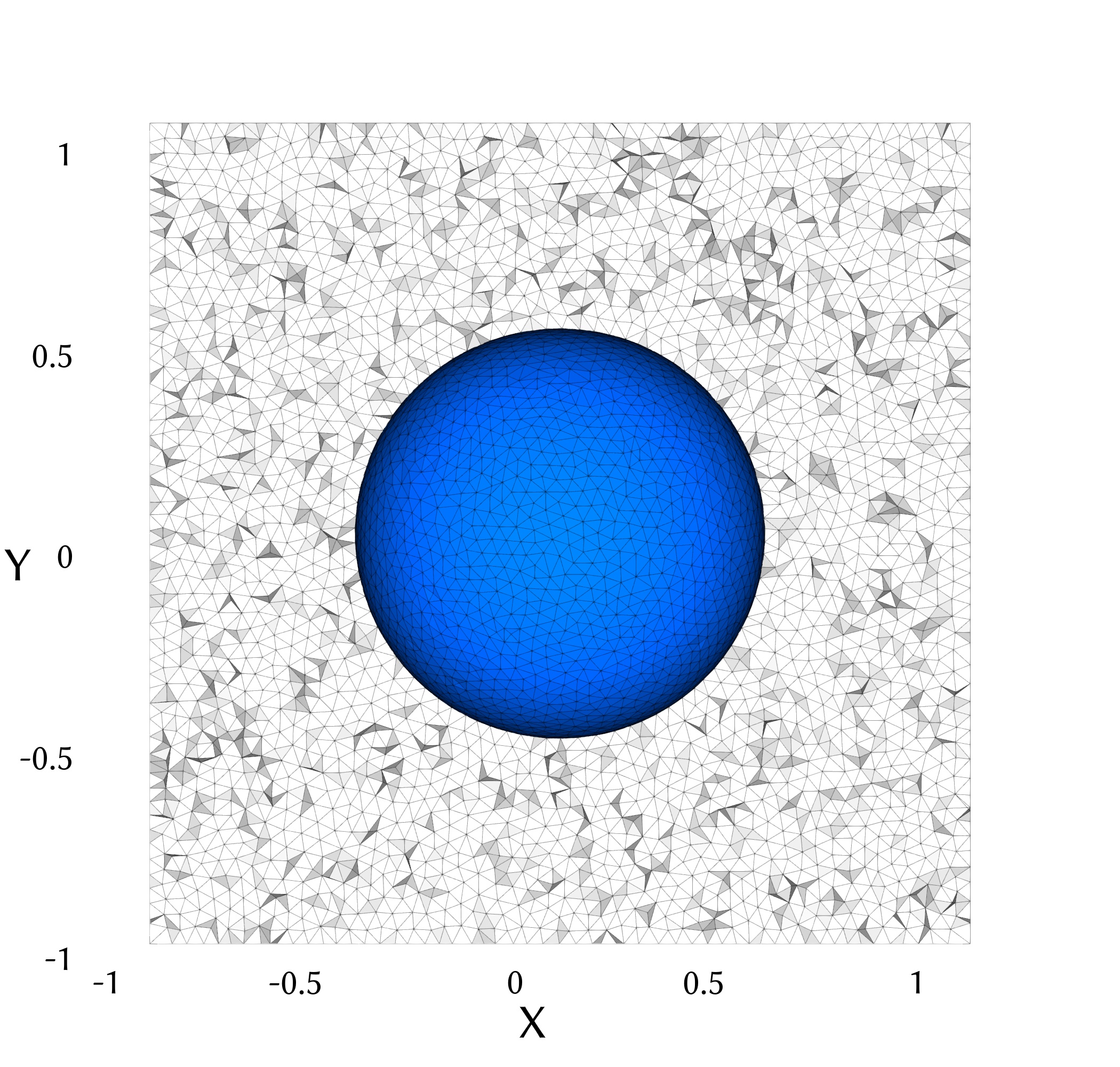}%
 }%
 \subcaptionbox{\label{subfig:3d:star:1} \(t=2.2\)}%
 [0.33\columnwidth]{%
 \includegraphics[width=0.33\columnwidth]{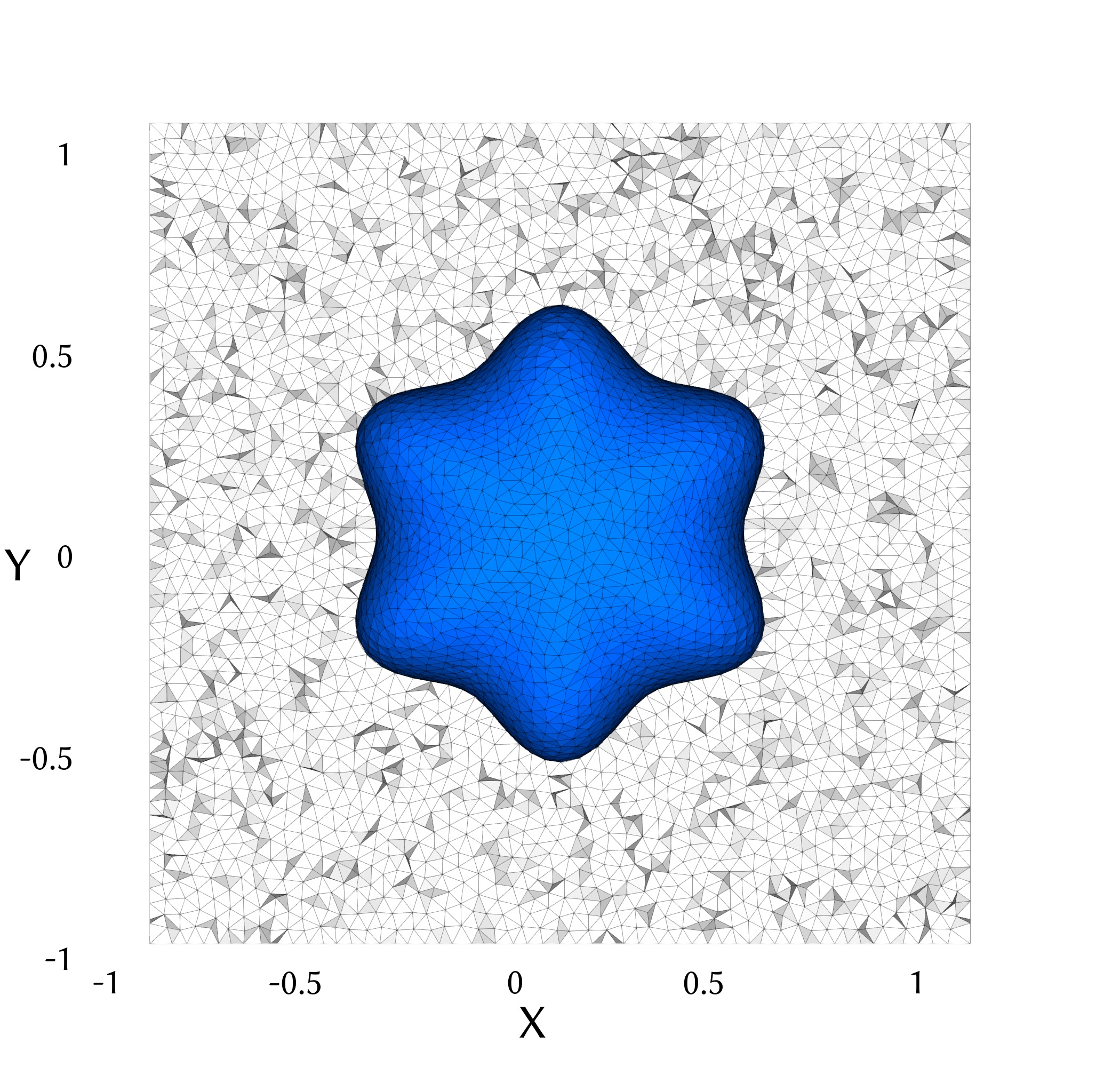}%
 }%
 \subcaptionbox{\label{subfig:3d:star:2} \(t=2.5\)}%
 [0.33\columnwidth]{%
 \includegraphics[width=0.33\columnwidth]{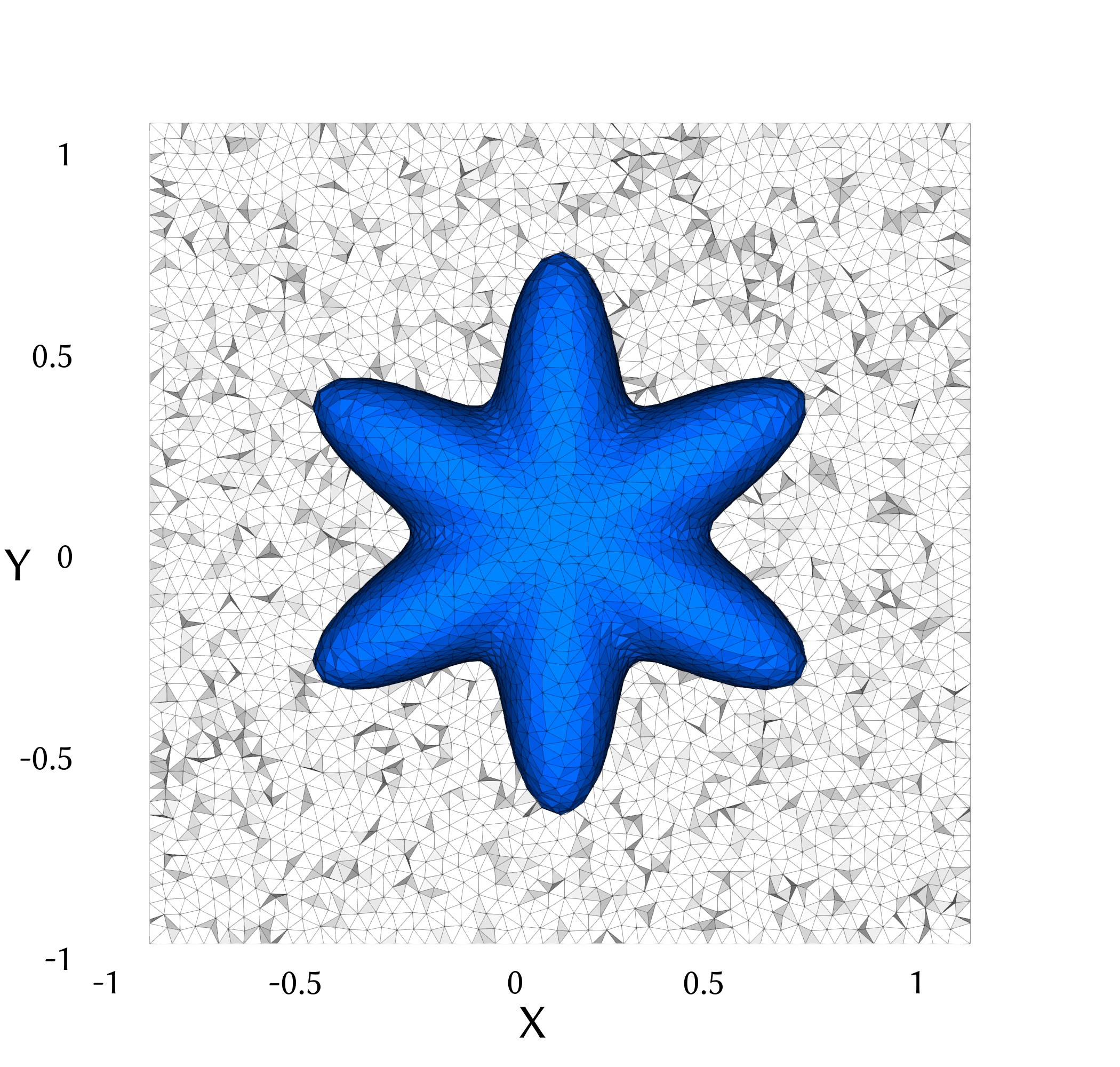}%
 }%
  \caption{%
    The moving interface (blue edges) for the star-shaped deformation example in three space dimensions.
    The plots show a slice through the computational domain at \(x_3 = 0\) of a mesh with in average  627000 cells. }
 \label{fig:mmesh_benchmark:star:3d}
\end{figure}

\begin{figure}[tp]
\centering
\includegraphics[height=0.33\linewidth]{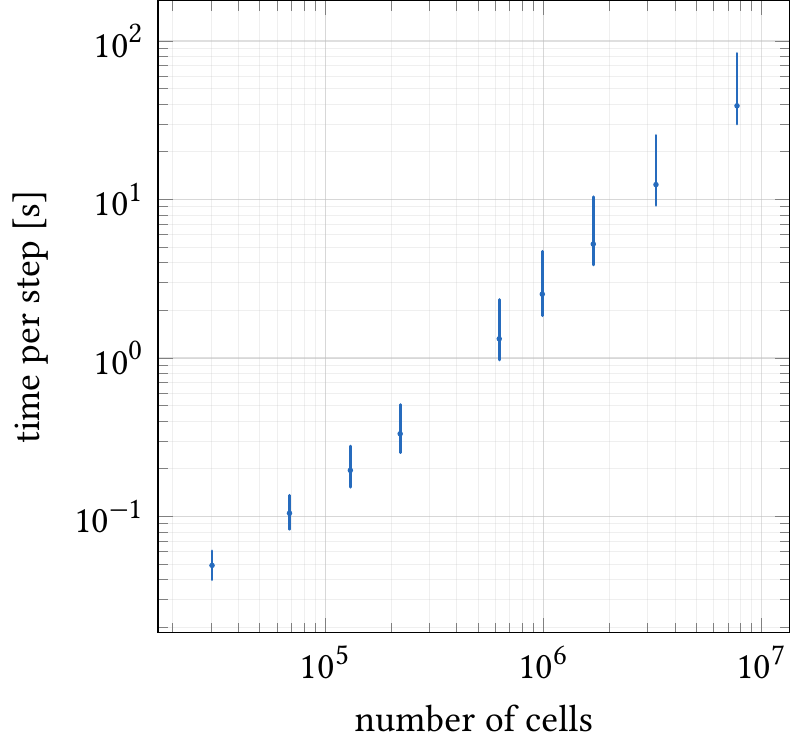}\hspace{0.1\columnwidth}\includegraphics[height=0.33\linewidth]{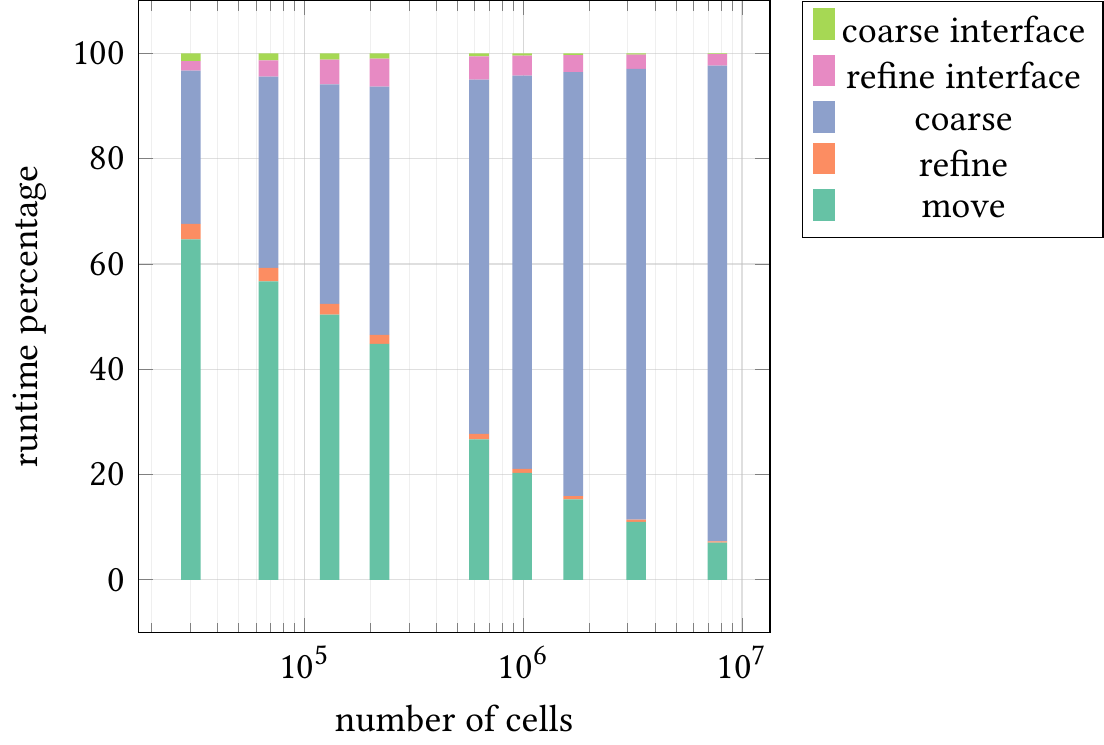}\caption{Computational time per step for the star-shaped deformation example in three space dimensions.
  In the left figure the lines represent the 90\%-percentile of the data over all time steps.
 The bars in the right figure indicate the mean computational time which is split into the respective mesh operation contributions:
move interface, refine/coarse (in Algorithm~\ref{alg:move_interface}), and refine/coarse interface.
 }
 \label{fig:3d:benchmark:star:runtime}
\end{figure}

We generalize the setup from the previous example to three space dimensions.
This time the moving interface \(\Gamma(t)\) starts as a sphere with radius \num{0.5} and center in the origin --- see Figure~\ref{fig:mmesh_benchmark:star:3d}~(\subref{subfig:3d:star:0}).
We consider the time interval from \(t_0 = 0\) to \(t_{\mathrm{end}} = 3\) with time steps $\Delta t = \num {2e-4}$.
The vector field describing the vertex motion is given by 
\begin{align}
  M &\colon [t_0, t_{\mathrm{end}}] \times \bR^3 \to \bR^3  \\
  &: (t, \vx) \mapsto \sin(2 \pi t) \Bigl(-x_1 \cos\bigl(2 \lceil t \rceil \operatorname{atan2}(x_2,x_1)\bigr), -x_2 \cos\bigl(2 \lceil t \rceil \operatorname{atan2}(x_2,x_1)\bigr), -x_3\Bigr).
\end{align}
Again, this results in a deformation of the sphere into a star-shaped object, as seen in Figure~\ref{fig:mmesh_benchmark:star:3d}~(\subref{subfig:3d:star:0})--(\subref{subfig:3d:star:2}).
\newline
We consider different mesh resolutions, where the initial mesh was generated with several maximum edge lengths \(\Delta x\). 
\newline
The computational time needed for each time step is shown in Figure~\ref{fig:3d:benchmark:star:runtime}.

We observe from Figure~\ref{fig:3d:benchmark:star:runtime} that again, as in the two-dimensional test case, the bulk coarsening will have asymptotically the largest proportion.
The computations at hand were done with the implementation given in \cite{alkaemper:interface:2021}, which serves as a comprehensible proof of concept.
Consequently, for runtime optimizations one should tackle the bulk coarsening method first and introduce more sophisticated data structures in order to achieve optimal runtime for the whole interface preserving moving mesh.

\subsection{Single Vortex Benchmark}

\begin{figure}[tp]
 \centering
 \subcaptionbox{\label{subfig:2d:vortex:0} \(t=0\)}
 [0.24\columnwidth]{
 \includegraphics[width=0.24\columnwidth]{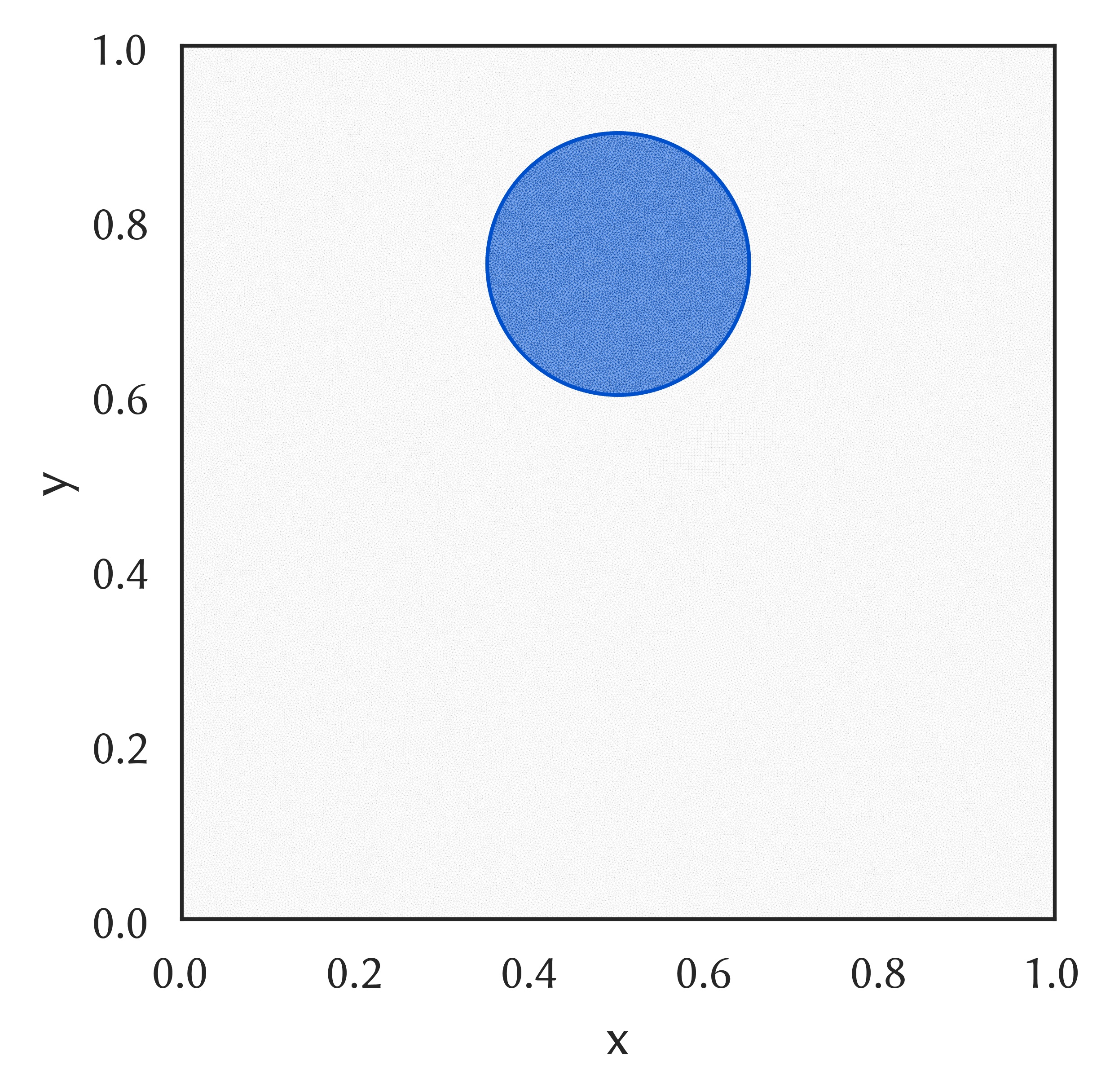}
 }
 \subcaptionbox{\label{subfig:2d:vortex:1} \(t=1\)}
 [0.24\columnwidth]{
 \includegraphics[width=0.24\columnwidth]{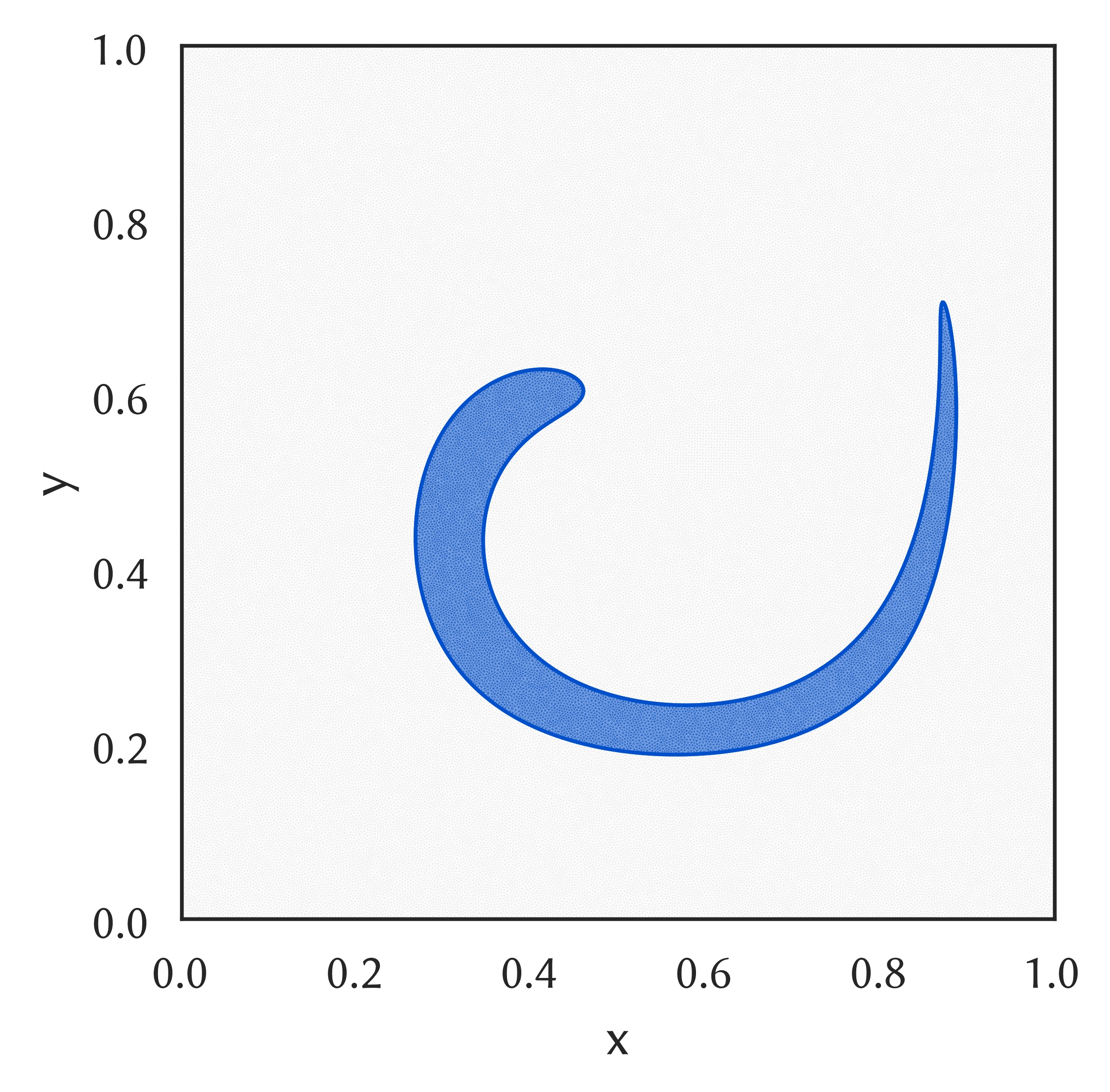}
 }
 \subcaptionbox{\label{subfig:2d:vortex:2} \(t=4\)}
 [0.24\columnwidth]{
 \includegraphics[width=0.24\columnwidth]{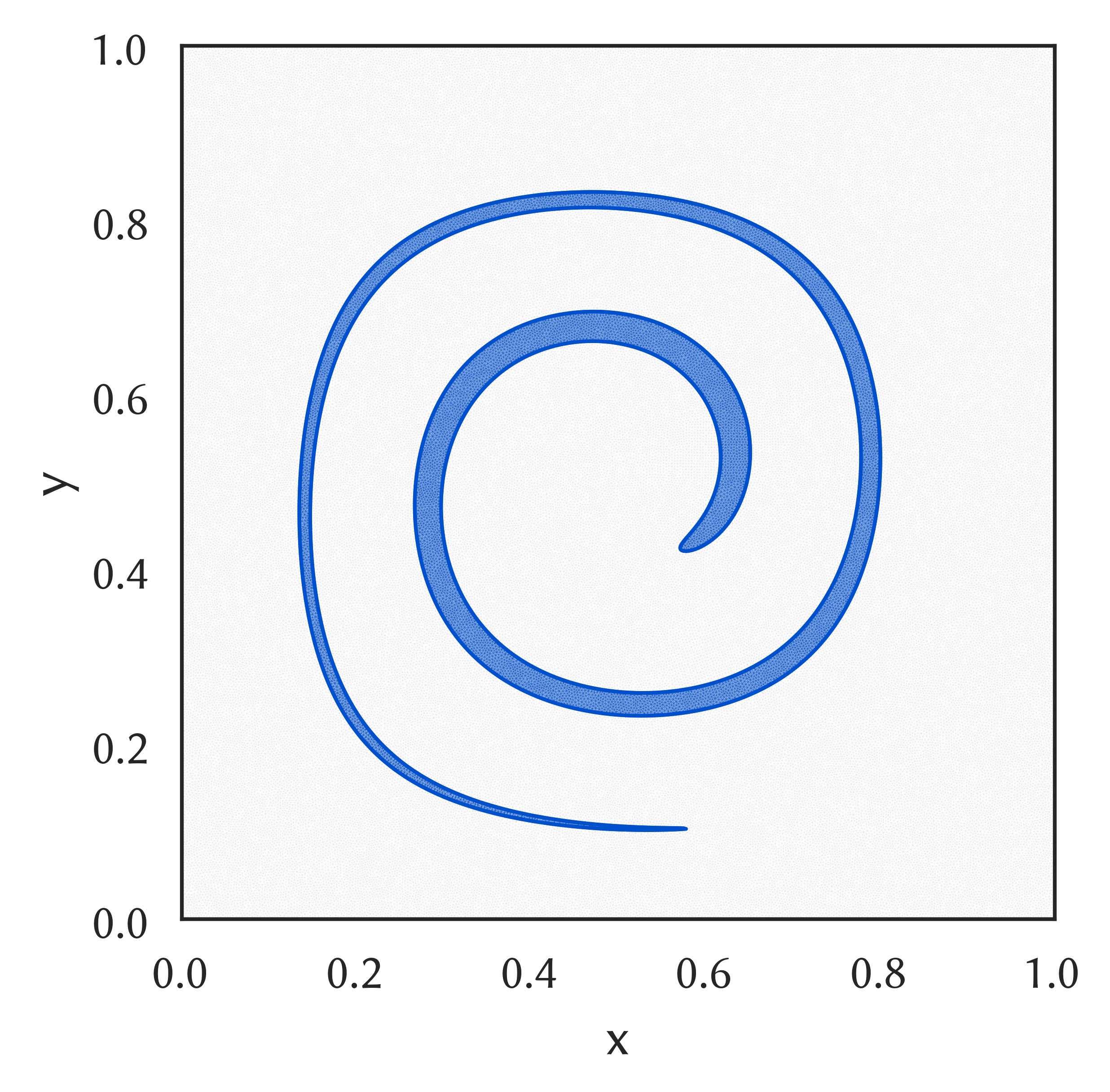}
 }
 \subcaptionbox{\label{subfig:2d:vortex:3} \(t=8\)}
 [0.24\columnwidth]{
 \includegraphics[width=0.24\columnwidth]{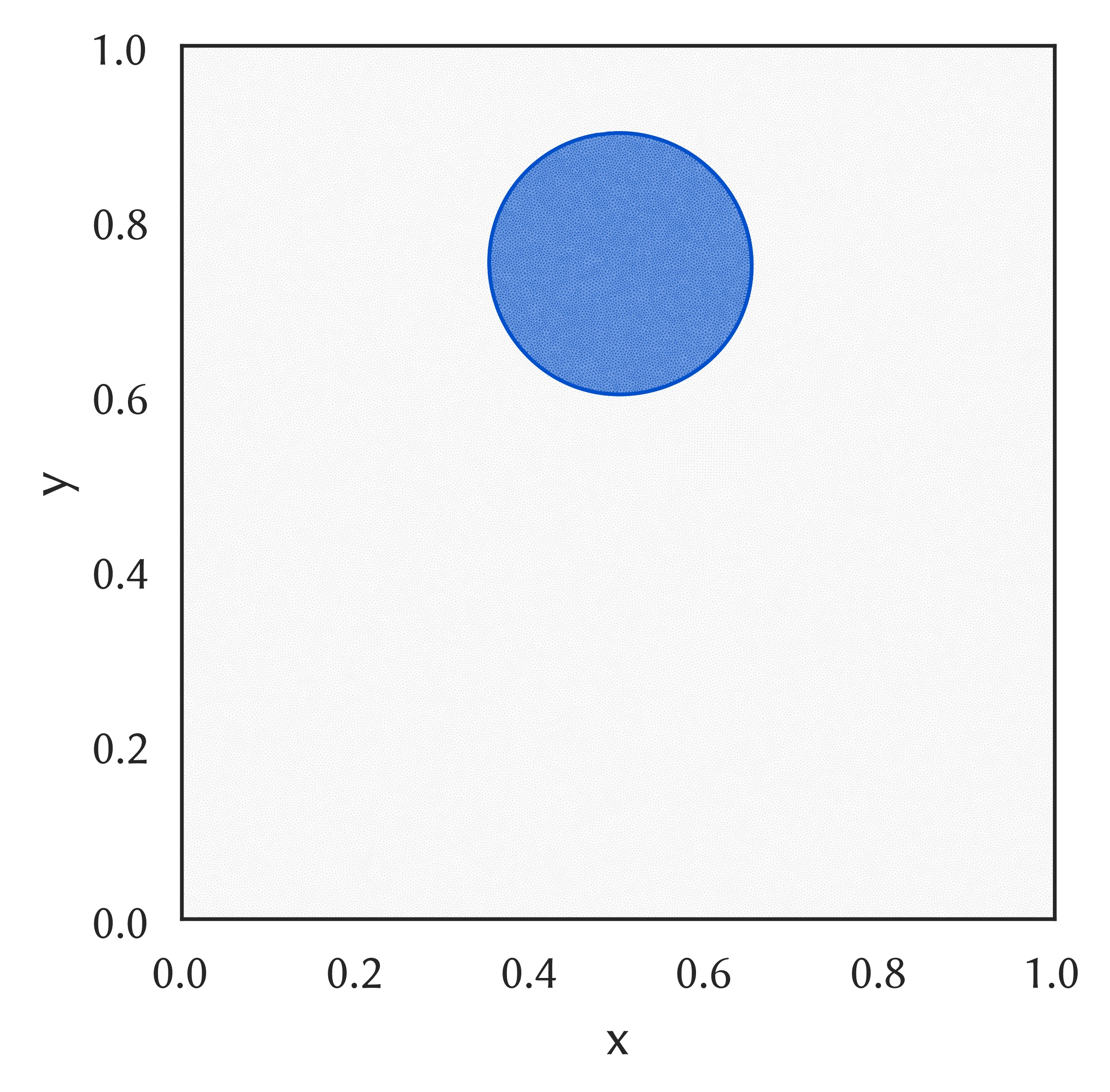}
 }
\caption{
    The moving interface (blue edges) for the vortex benchmark in two space dimensions.
    The depicted example was computed on a mesh with in average 202000 cells. }
 \label{fig:mmesh_benchmark:vortex:2d}
\end{figure}

\begin{figure}[tp]
 \centering
 \subcaptionbox{\label{subfig:3d:vortex:0} \(t=0\)}
 [0.24\columnwidth]{
 \includegraphics[width=0.24\columnwidth]{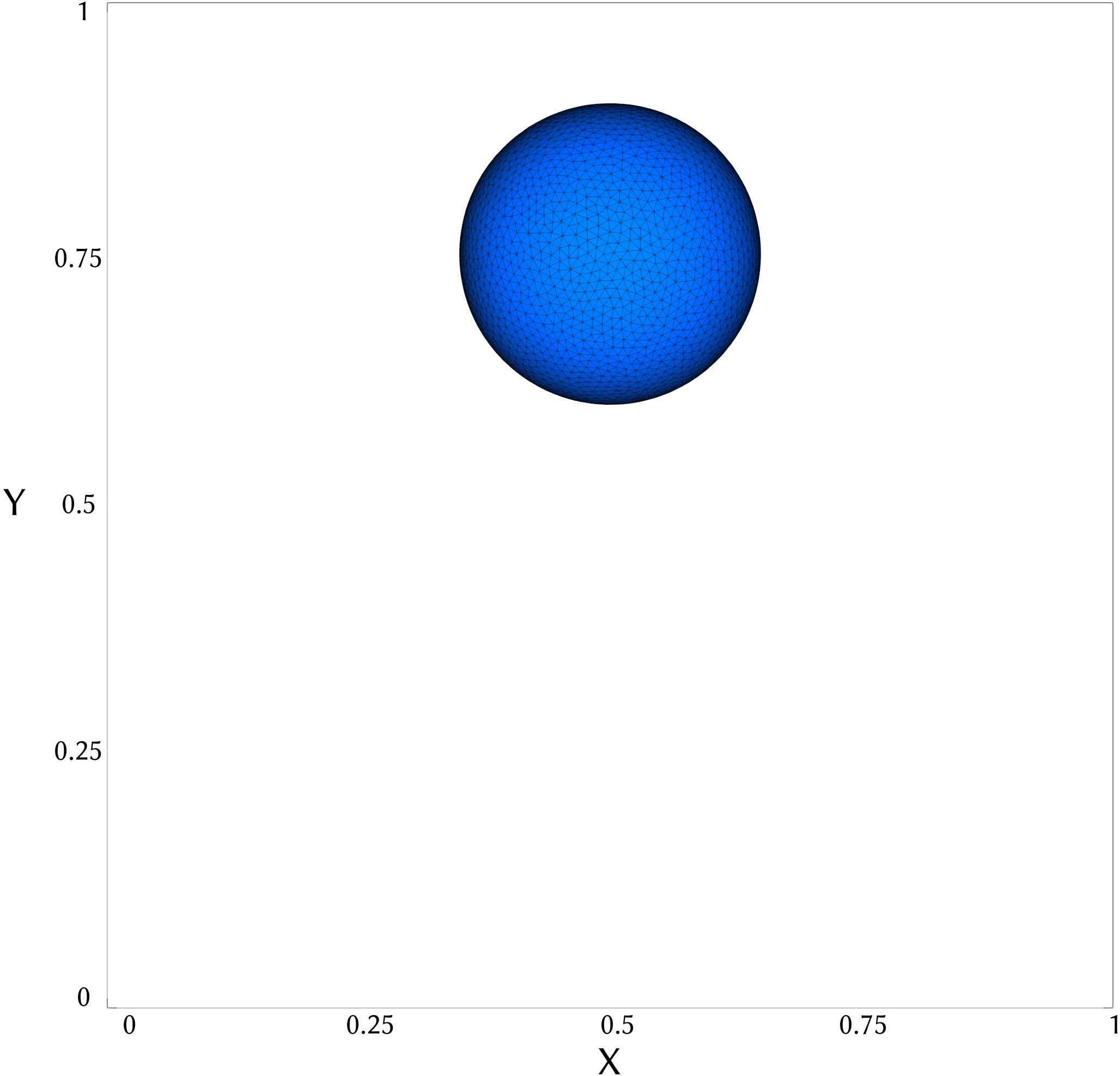}
 }
 \subcaptionbox{\label{subfig:3d:vortex:1} \(t=1\)}
 [0.24\columnwidth]{
 \includegraphics[width=0.24\columnwidth]{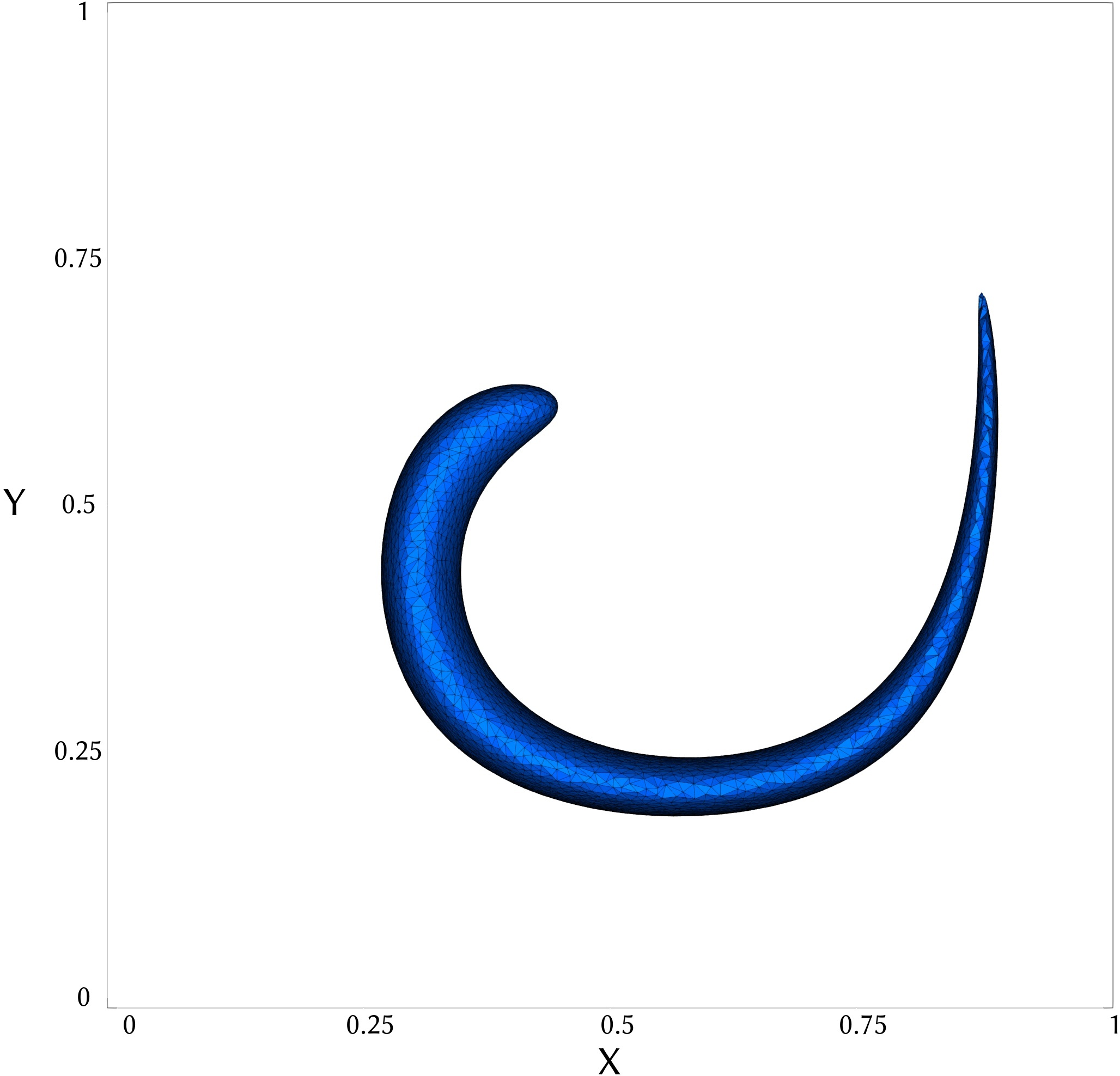}
 }
 \subcaptionbox{\label{subfig:3d:vortex:2} \(t=3\)}
 [0.24\columnwidth]{
 \includegraphics[width=0.24\columnwidth]{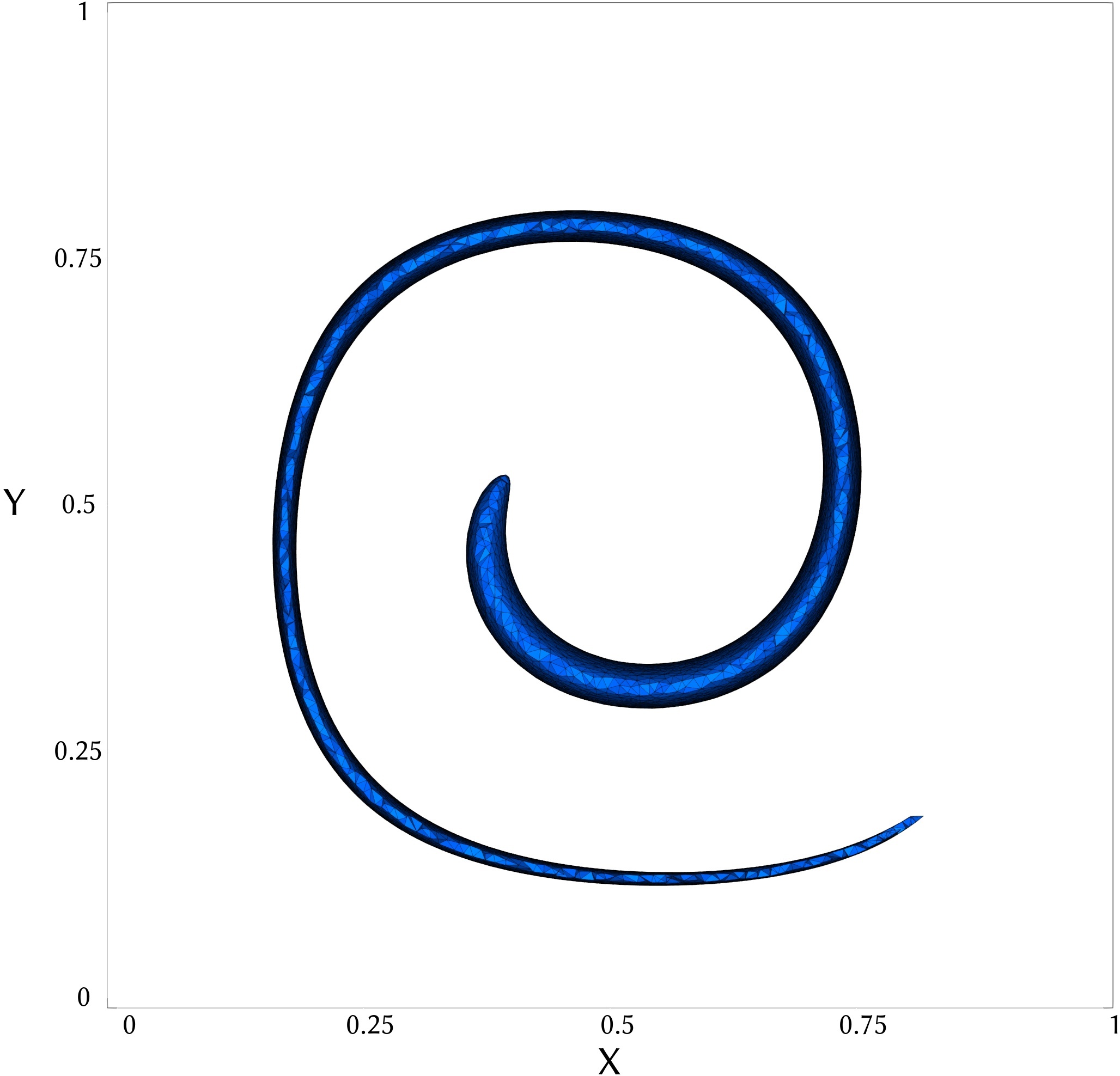}
 }
 \subcaptionbox{\label{subfig:3d:vortex:3} \(t=6\)}
 [0.24\columnwidth]{
 \includegraphics[width=0.24\columnwidth]{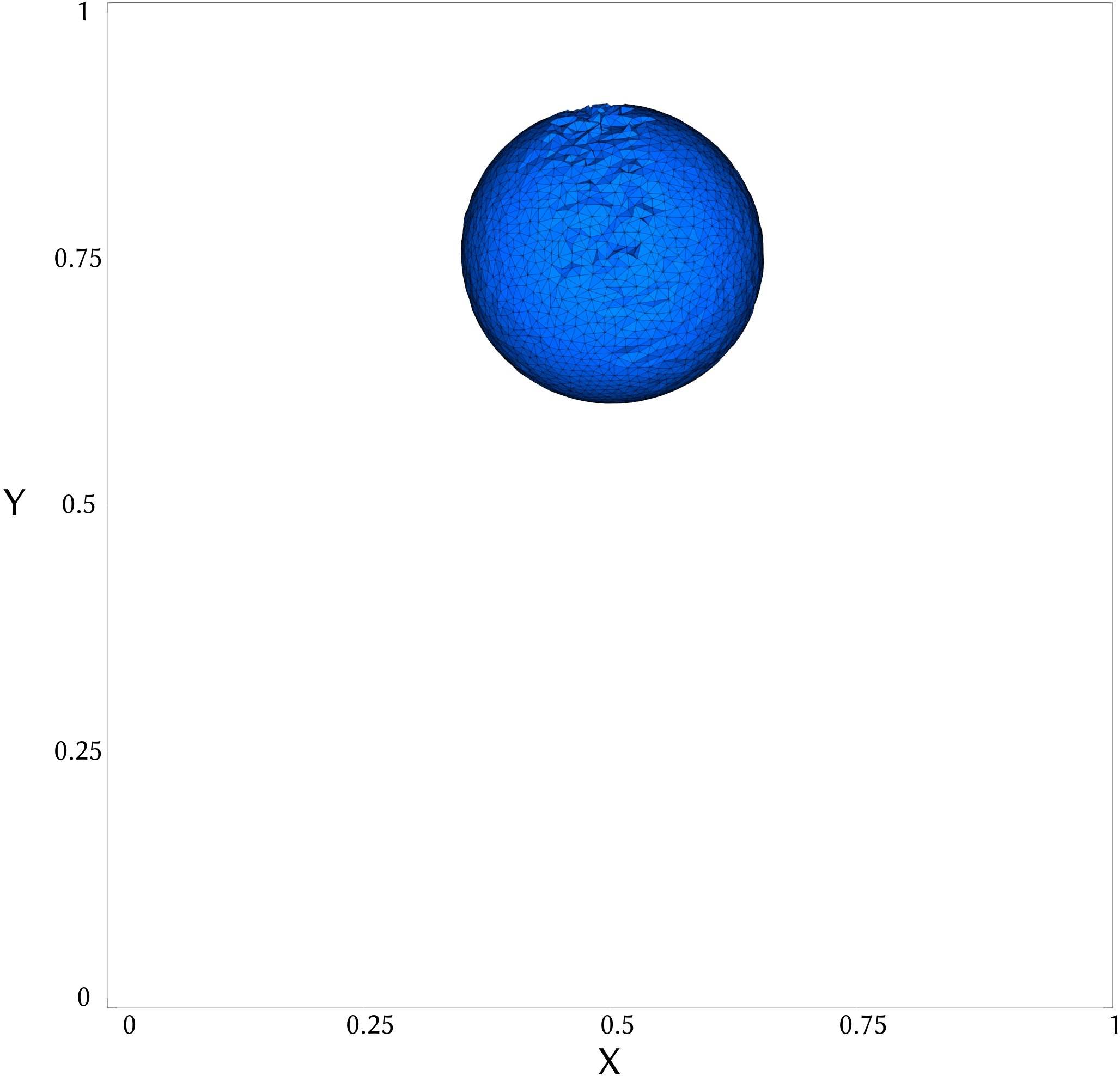}
 }
\caption{
    The moving interface (blue surface) for the vortex benchmark in three space dimensions.
    The depicted example was computed on a mesh with in average 2191000 cells. }
 \label{fig:mmesh_benchmark:vortex:3d}
\end{figure}

A benchmark problem, which is widely used in the context of Volume-of-Fluid (VoF) methods with surface reconstruction, is the deformation of a sphere into a single vortex by a prescribed velocity field, see e.g. \cite{rider.kothe:1998}.

\subsubsection{The Two-Dimensional Algorithm}

In two space dimension the single vortex is induced by the velocity field
\begin{align*}
   M &\colon 
  [t_0, t_{\mathrm{end}}] \times \bR^2 \to \bR^2 \\
  &: 
  (t,\vx) \mapsto \cos\bigl(\tfrac{\pi t}{2 t_{\mathrm{end}}}\bigr) \Bigl(2\cos(\pi x_1)\sin(\pi x_1)\sin^2(\pi x_2), 2\cos(\pi x_2)\sin(\pi x_2)\sin^2(\pi x_1)\Bigr)
\end{align*}
with $t_{\mathrm{end}} = 8$.
The moving interface \(\Gamma(t)\) is initially a circle with radius \num{0.15} and center at \((0.5, 0.75)\) inside the domain \([0,1]^2\).
The velocity field transforms the initial circle to a spiral shape.
The factor \(\cos(\tfrac{\pi t}{2 t_{\mathrm{end}}})\) reverts the velocity field at time \(t=4\) such that the interface returns to its initial circular shape by the time $t_{\mathrm{end}} = 8$.
The results computed with the time step $\Delta t = \num{e-4}$ are shown in Figure~\ref{fig:mmesh_benchmark:vortex:2d}~(\subref{subfig:2d:vortex:0})--(\subref{subfig:2d:vortex:3}).
In contrast to VoF methods with reconstruction no breakup of the interface occurs by construction of the interface preserving moving mesh and the interface returns into almost perfect circular shape.
The enclosed area at initial time $t_0$ and at final time $t_{\mathrm{end}}$ is \num{7.068e-2} and \num{7.070e-2}, which gives an absolute error of \num{3.352e-05} and \num{2.489e-4} in the area, respectively.
The signed deviation of all interface vertices from the circle is in a range from \num{-2.720e-3} to \num{3.141e-3} with a mean deviation of \num{3.592e-4}.

\subsubsection{The Three-Dimensional Algorithm}

Similarly, in the three-dimensional setting the single vortex is induced by the velocity field
\begin{align*}
   M &\colon 
  [t_0, t_{\mathrm{end}}] \times \bR^3 \to \bR^3 \\
  &: 
  (t,\vx) \mapsto \cos\bigl(\tfrac{\pi t}{2 t_{\mathrm{end}}}\bigr) \Bigl(2\cos(\pi x_1)\sin(\pi x_1)\sin^2(\pi x_2), 2\cos(\pi x_2)\sin(\pi x_2)\sin^2(\pi x_1), 0\Bigr)
\end{align*}
with $t_{\mathrm{end}} = 6$.
Again, the moving interface \(\Gamma(t)\) is initially a sphere with radius \num{0.15} and center at \((0.5, 0.75, 0)\) inside the domain \([-1,1]^2 \times [-0.5,0.5]\).
The velocity field transforms the initial sphere to a spiral shape by the time \(t=3\) and transforms it back to its initial spherical shape by the time $t_{\mathrm{end}} = 6$.
The resulting moving interface computed with the time step $\Delta t = \num{2e-4}$ is depicted in Figure~\ref{fig:mmesh_benchmark:vortex:3d}~(\subref{subfig:3d:vortex:0})--(\subref{subfig:3d:vortex:3}).
Unlike in the two-dimensional case the strong deformation of the sphere causes changes of the interface surface.
More precisely, interface facets may flip on the interface, which gives in combination with the interface remeshing a less precise resolution of the interface surface for the three-dimensional setting.
Nevertheless, the final state is still in good agreement with the predicted spherical shape.
The enclosed volume at initial time $t_0$ and at final time $t_{\mathrm{end}}$ is \num{1.408e-2} and \num{1.380e-2}, which gives an absolute error of \num{4.108e-3} and \num{2.373e-2} in the volume, respectively.
The signed deviation of all interface vertices from the sphere is in a range from \num{-1.196e-2} to \num{3.473e-3} with a mean deviation of \num{-1.044e-3}.

In summary this test case confirms that with the IPMM even extreme interface deformations are possible.
The interface surface is directly resolved within the full-dimensional mesh, which bypasses surface reconstruction steps and easily enables the computation of curvature, e.g. by least-squares fitting spheres.
Moreover, the interface surfaces stay connected for all times both in two and three space dimensions.

\subsection{Circular Advection Example}
\label{subsec:mmesh_benchmark:circular}

In the previous examples we presented some calculations, where the motion of the interface was given explicitly.
In this final example we will treat the case, where the motion of the interface depends dynamically on the numerical solution of an (advective) PDE on the full dimensional mesh.
As such, the numerical data on the surrounding full-dimensional mesh interacts directly with the interface motion.

\subsubsection{The Two-Dimensional Algorithm}

We regard the circular advection around the origin in two space dimensions.
The model problem reads
\begin{align} 
\label{eq:circadv}
\begin{array}{r c l l}
\frac{\dd}{\dd t}\left( u(t,x) \right) + \frac{\dd}{\dd x_1}\left( -x_2 u(t,x) \right) + \frac{\dd}{\dd x_2}\left( x_1 u(t,x) \right) & = & 0, & \text{ for } x\in \Omega, t\in (0,t_{\mathrm{end}}),\\
u(0,x) & = & u_0(x), & \text{ for } x \in \Omega.
\end{array}
\end{align}

\begin{figure}[tp]
 \centering
\includegraphics[height=0.30\columnwidth]{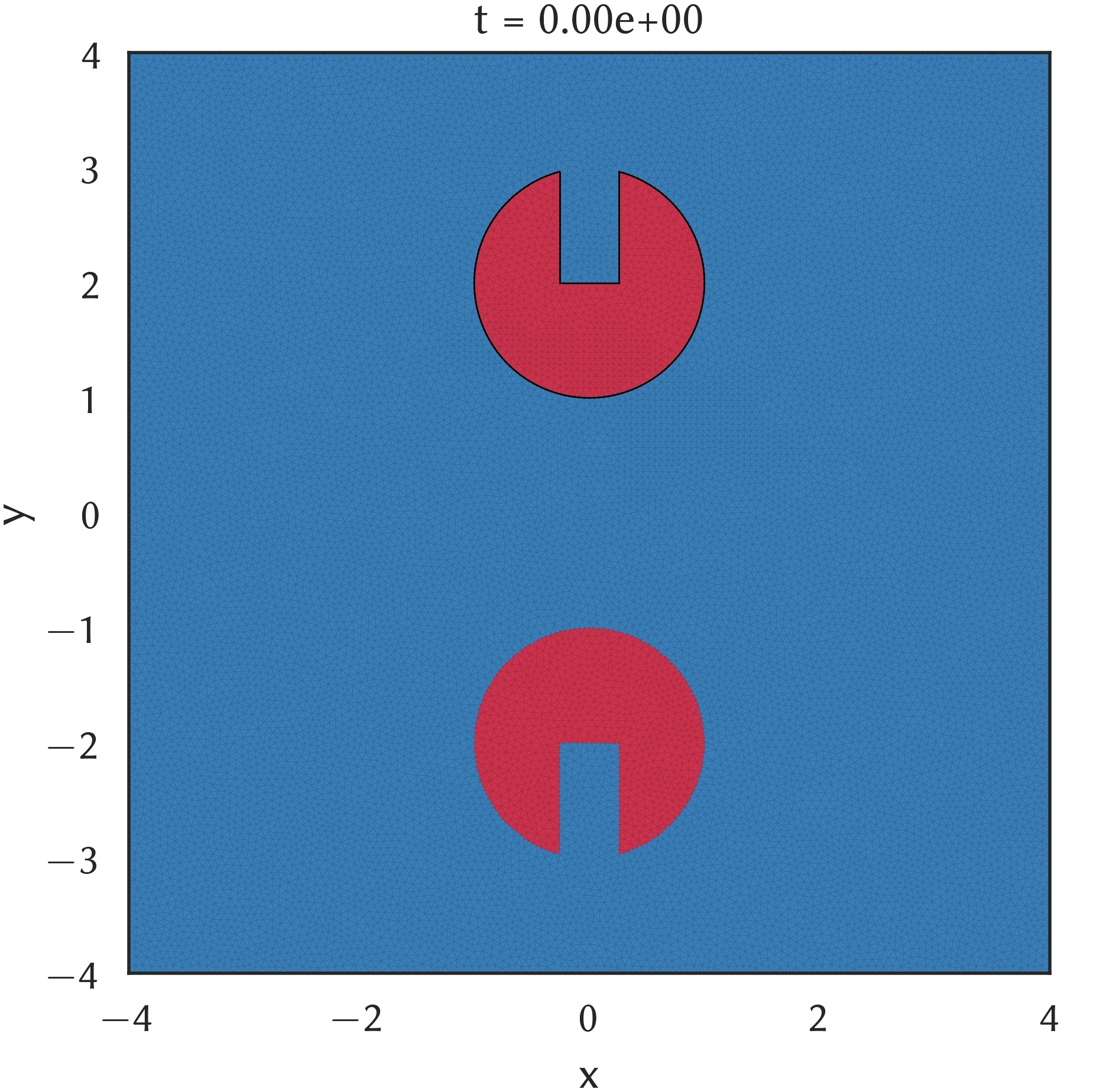}
\includegraphics[height=0.30\columnwidth]{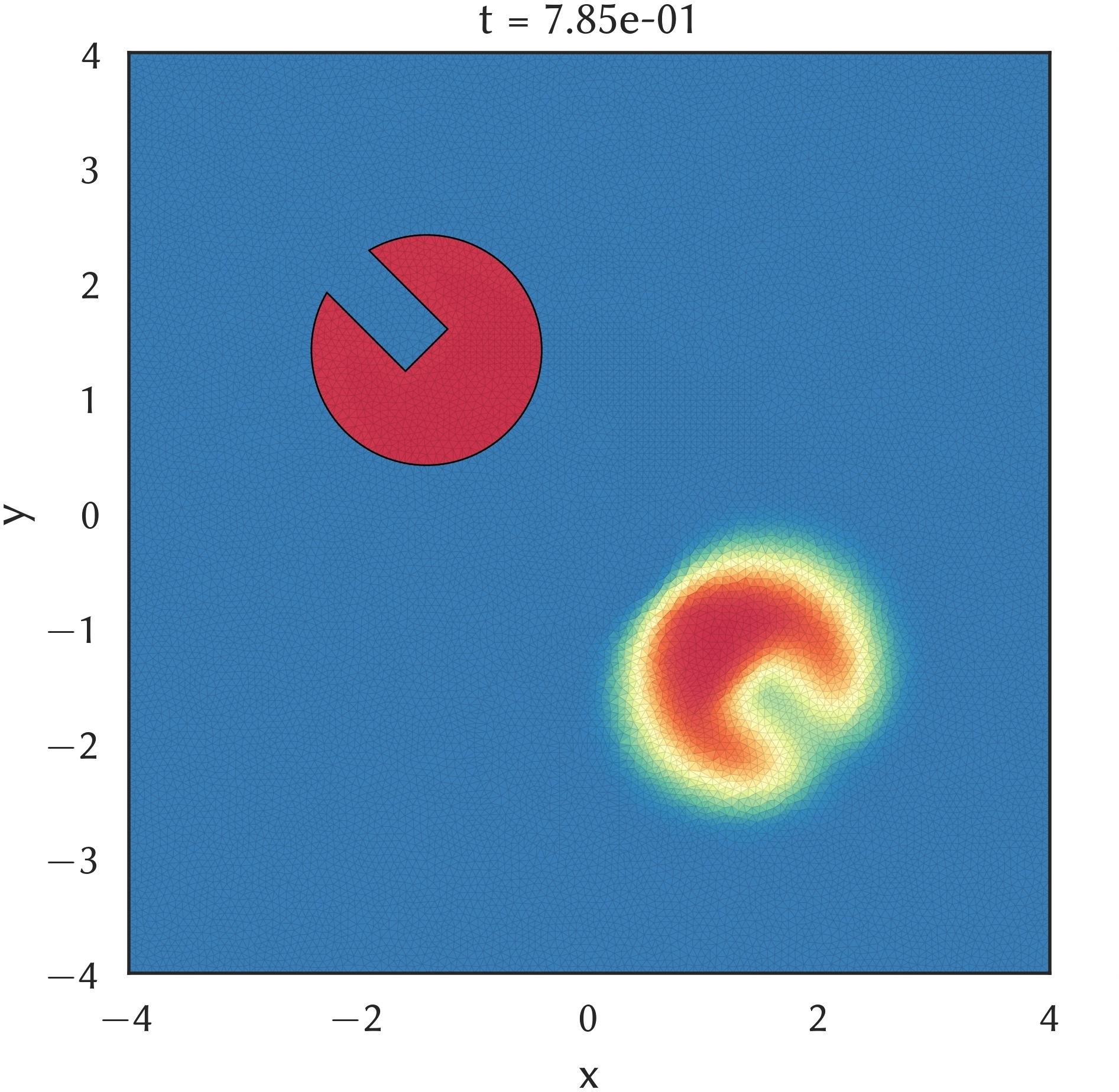}
\includegraphics[height=0.30\columnwidth]{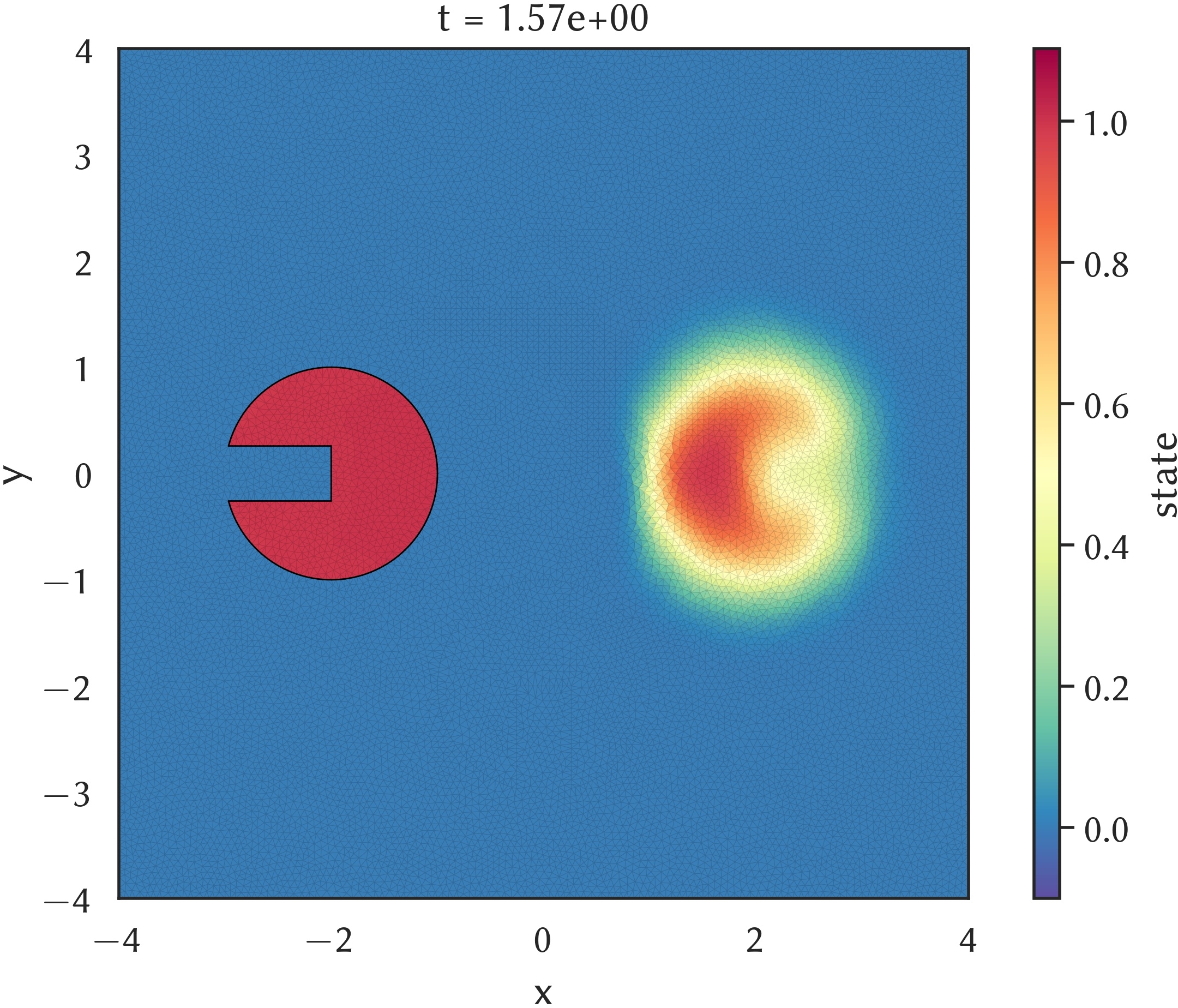}
\caption{
 Two sliced disks that are rotated via circular advection.
The initially upper disk is treated with the IPMM-FV method whereas the initially lower disk is treated with the standard FV method.
The first picture shows the initial state in a mesh with 32708 cells and mesh edge lengths in a range from \num{0.0426} to \num{0.1266}. The second picture shows an intermediate state at $t=\pi/4$. The third picture shows the final state at $t_{\mathrm{end}}=\pi/2$ after a quarter rotation.
For the computation a fixed time step \(\Delta t = \num{e-4}\) was used.
 }
 \label{fig:2d:benchmark_two_sliced_disks}
\end{figure}

The exact solution is a counterclockwise rotation around the origin of the initial data $u_0$.
As initial state we choose an indicator function on a sliced disk domain, \(\Omega=[-4,4]^2\) and $t_{\mathrm{end}}=\pi/2$, see also Figure~\ref{fig:2d:benchmark_two_sliced_disks}.
The setting is chosen similar to the seminal Zalesak benchmark problem \cite{zalesak:1979}.

To compute the cell data, as well as the interface motion, we will make use of the Finite Volume method as e.g. in \cite{chalons.rohde.ea:finite:2017} on top of the interface preserving moving mesh, which we will call interface preserving moving mesh finite volume method (IPMM-FV method).

In Table~\ref{tab:benchmark_two_sliced_disks:2d} we list both runtime and numerical error for the standard finite volume (FV) method versus the IPMM-FV method.
One can see from the table, that the $L^1$-error of the IPMM-FV method is very close to the initial $L^1$-error, which results from the piecewise linear approximation of the interface.
The FV method takes approximately half of the runtime of the IPMM-FV method on meshes with similar discretization widths.
Nevertheless, even the FV method on the finest mesh does not reach the accuracy of the IPMM-FV method on the coarsest mesh.

\begin{table}[tp]
\centering
\resizebox{\columnwidth}{!}{%
\begin{tabular}{c|lll|lll}
  \toprule
& \multicolumn{3}{c|}{{\bfseries IPMM-FV method}{\bfseries -- 2d}} & \multicolumn{3}{c}{{\bfseries FV method}{\bfseries -- 2d}} \\
  initial $L^1$-error &                                  mesh width &         $L^1$-error &       time/step [s] &                             mesh width &        $L^1$-error &       time/step [s] \\
  \midrule
  $4.9 \cdot 10^{-2}$ &                         $5.2 \cdot 10^{-1}$ & $4.8 \cdot 10^{-2}$ & $6.6 \cdot 10^{-4}$ &                    $5.2 \cdot 10^{-1}$ & $3.1 \cdot 10^{0}$ & $2.8 \cdot 10^{-4}$ \\
  $2.4 \cdot 10^{-2}$ &                         $3.9 \cdot 10^{-1}$ & $2.3 \cdot 10^{-2}$ & $1.1 \cdot 10^{-3}$ &                    $3.9 \cdot 10^{-1}$ & $2.8 \cdot 10^{0}$ & $5.3 \cdot 10^{-4}$ \\
  $7.2 \cdot 10^{-3}$ &                         $2.3 \cdot 10^{-1}$ & $8.1 \cdot 10^{-3}$ & $2.9 \cdot 10^{-3}$ &                    $2.3 \cdot 10^{-1}$ & $2.4 \cdot 10^{0}$ & $1.6 \cdot 10^{-3}$ \\
  $3.8 \cdot 10^{-3}$ &                         $1.5 \cdot 10^{-1}$ & $3.6 \cdot 10^{-3}$ & $6.6 \cdot 10^{-3}$ &                    $1.5 \cdot 10^{-1}$ & $2.1 \cdot 10^{0}$ & $4.1 \cdot 10^{-3}$ \\
  $2.0 \cdot 10^{-3}$ &                         $1.1 \cdot 10^{-1}$ & $2.1 \cdot 10^{-3}$ & $1.3 \cdot 10^{-2}$ &                    $1.1 \cdot 10^{-1}$ & $1.9 \cdot 10^{0}$ & $7.9 \cdot 10^{-3}$ \\
  $9.7 \cdot 10^{-4}$ &                         $7.5 \cdot 10^{-2}$ & $1.4 \cdot 10^{-3}$ & $3.3 \cdot 10^{-2}$ &                    $7.5 \cdot 10^{-2}$ & $1.6 \cdot 10^{0}$ & $2.5 \cdot 10^{-2}$ \\
  $2.0 \cdot 10^{-4}$ &                         $3.8 \cdot 10^{-2}$ & $8.8 \cdot 10^{-4}$ & $2.9 \cdot 10^{-1}$ &                    $3.8 \cdot 10^{-2}$ & $1.2 \cdot 10^{0}$ & $2.4 \cdot 10^{-1}$ \\
  \bottomrule
  \end{tabular}
  }%
  \caption{Comparison of the  $L^1$-error at time $t_{\mathrm{end}}=\pi/2$ and runtime of the IPMM-FV method and of the FV method for the sliced disk benchmark case computed with a fixed time step of $\Delta t = \num{e-4}$ in two space dimensions.
The mesh widths are the average edge lengths that appear during the simulation.}
\label{tab:benchmark_two_sliced_disks:2d}
\end{table}

\subsubsection{The Three-Dimensional Algorithm}

By using the same set of equations \eqref{eq:circadv} we can define the circular advection around the $z$-axis in three space dimensions.

We generalize the previous setting by choosing an indicator function on a sliced sphere as initial state and \(\Omega=[-4,4]^2\times[-2,2]\).

Again, we list both runtime and numerical error for the standard finite volume (FV) method versus the IPMM-FV method, see Table~\ref{tab:benchmark_two_sliced_disks:3d}.
In the three-dimensional setting the IPMM-FV method takes between 2 to 12 times the runtime of the FV method on meshes with similar discretization widths.
The finer the mesh is chosen the closer the runtime gets for both methods.
This indicates that the runtime complexity is similar to the one needed for the numerical operations.
For more complex applications the runtime for the IPMM may even become negligible in comparsion to further numerical computations.
Nevertheless, we see that also in three space dimensions the FV method does not reach the accuracy of the IPMM-FV method.

In both compared methods a standard finite volume method in the bulk regions was used.
Of course the interface preserving moving mesh is not limited to finite volume methods only but can also be used in combination with higher order Galerkin methods, which is a subject of ongoing research.

\begin{table}[tp]
  \centering
\resizebox{\columnwidth}{!}{%
\begin{tabular}{c|lll|lll}
  \toprule
  & \multicolumn{3}{c|}{{\bfseries IPMM-FV method}{\bfseries -- 3d}} & \multicolumn{3}{c}{{\bfseries FV method}{\bfseries -- 3d}} \\
  initial $L^1$-error &                                  mesh width &         $L^1$-error &       time/step [s] &                             mesh width &        $L^1$-error &       time/step [s] \\
  \midrule
  $2.0 \cdot 10^{-1}$ &                          $1.3 \cdot 10^{0}$ & $2.6 \cdot 10^{-1}$ & $2.4 \cdot 10^{-2}$ &                     $1.4 \cdot 10^{0}$ & $5.2 \cdot 10^{0}$ & $2.0 \cdot 10^{-3}$ \\
  $1.9 \cdot 10^{-1}$ &                          $1.2 \cdot 10^{0}$ & $2.6 \cdot 10^{-1}$ & $2.6 \cdot 10^{-2}$ &                     $1.3 \cdot 10^{0}$ & $5.1 \cdot 10^{0}$ & $2.4 \cdot 10^{-3}$ \\
  $1.9 \cdot 10^{-1}$ &                          $1.1 \cdot 10^{0}$ & $2.6 \cdot 10^{-1}$ & $2.6 \cdot 10^{-2}$ &                     $1.1 \cdot 10^{0}$ & $4.9 \cdot 10^{0}$ & $3.2 \cdot 10^{-3}$ \\
  $1.9 \cdot 10^{-1}$ &                         $8.5 \cdot 10^{-1}$ & $2.6 \cdot 10^{-1}$ & $3.0 \cdot 10^{-2}$ &                    $8.6 \cdot 10^{-1}$ & $4.7 \cdot 10^{0}$ & $5.5 \cdot 10^{-3}$ \\
  $1.6 \cdot 10^{-1}$ &                         $5.9 \cdot 10^{-1}$ & $2.3 \cdot 10^{-1}$ & $4.6 \cdot 10^{-2}$ &                    $5.8 \cdot 10^{-1}$ & $4.3 \cdot 10^{0}$ & $1.5 \cdot 10^{-2}$ \\
  $1.2 \cdot 10^{-1}$ &                         $4.4 \cdot 10^{-1}$ & $1.9 \cdot 10^{-1}$ & $8.1 \cdot 10^{-2}$ &                    $4.4 \cdot 10^{-1}$ & $3.9 \cdot 10^{0}$ & $3.8 \cdot 10^{-2}$ \\
  $5.8 \cdot 10^{-2}$ &                         $2.9 \cdot 10^{-1}$ & $1.2 \cdot 10^{-1}$ & $2.6 \cdot 10^{-1}$ &                    $2.9 \cdot 10^{-1}$ & $3.5 \cdot 10^{0}$ & $1.6 \cdot 10^{-1}$ \\
  $3.0 \cdot 10^{-2}$ &                         $2.2 \cdot 10^{-1}$ & $9.5 \cdot 10^{-2}$ & $6.5 \cdot 10^{-1}$ &                    $2.2 \cdot 10^{-1}$ & $3.2 \cdot 10^{0}$ & $3.9 \cdot 10^{-1}$ \\
  $7.6 \cdot 10^{-3}$ &                         $1.5 \cdot 10^{-1}$ & $7.4 \cdot 10^{-2}$ &  $2.6 \cdot 10^{0}$ &                    $1.5 \cdot 10^{-1}$ & $2.8 \cdot 10^{0}$ &  $1.4 \cdot 10^{0}$ \\
  \bottomrule
  \end{tabular}
}%
  \caption{Comparison of the $L^1$-error at time $t_{\mathrm{end}}=\pi/2$ and runtime of the IPMM-FV method and of the FV method for the sliced sphere benchmark case computed with a fixed time step of $\Delta t = \num{e-4}$ in three space dimensions.
The mesh widths are the average edge lengths that appear during the simulation.}
\label{tab:benchmark_two_sliced_disks:3d}
\end{table}

\section{Conclusions}
Moving mesh methods are a key tool to enable the robust numerical solution of PDE-based free boundary value problems.
We introduced in this contribution a new moving mesh (method) for tracking the motion of the free boundary interface as a codimension-$1$ manifold in a meshed bulk domain such that the  manifold is represented for all times as a subset of facets of the bulk mesh.
Whilst the approach works for arbitrary bulk dimensions, an open source implementation for the two- and three-dimensional case can be found at \cite{alkaemper:interface:2021}.
Remeshing by interface coarsening and refinement is integrated in the entire algorithm preventing the formation of numerically disadvantageous cells and thus ensuring a desired mesh resolution.
Notably, the remeshing operations are performed in a local manner which is even shown rigorously in Theorem~\ref{thm:minsphere} for the interface coarsening algorithm.
In this way our approach is radically different from the widely used Lagrangian approaches, see e.g. Figure~\ref{fig:topology_changes_mesh} as well as the  interface-reconstruction approaches (e.g. piecewise linear interface construction (PLIC) in the context of VoF methods or  cut-cell approaches).
The performance of the IPMM is confirmed by the results of the test cases in Section~\ref{sec:mmesh_benchmark}.
Moreover, we show that it is capable to resolve  strong interface deformations and large changes of the  interface geometry, while always preserving the $(d-1)$-dimensional interface within the mesh.
We provide evidence that the runtime complexity per time step is similar to the one needed for the numerical operations.
The IPMM method is easily adapted to modern solver techniques for PDE-based problems like high-order finite element or discontinuous Galerkin methods.
To be consistent with high-order numerics the geometrical error of the interface position should scale in the same way leading to isoparametric representations of the interface facets.
Future work will aim at exploiting the local nature of the algorithms in order to run the algorithms in parallel on a distributed mesh.
Since dynamic (bulk) mesh adaption became a standard tool in the last decade adaptive control of the moving mesh becomes desirable.
Finally, our moving mesh propels the design of conforming numerical discretization methods for free boundary problems with inherent surface dynamics models like Marangoni convection.

\subsection*{Acknowledgements}
\label{sec:acknowledgements}
This work was supported by the Deutsche Forschungsgemeinschaft (DFG, German Research Foundation) through the project  \mbox{SFB--TRR 75} ``Droplet dynamics under extreme ambient conditions'' with the project number 84292822 
and the DFG under Germany's Excellence Strategy - EXC 2075 with the project number 390740016.

\printbibliography

\end{document}